\documentclass[%
reprint,
superscriptaddress,
amsmath,amssymb,
aps,
pra,
]{revtex4-2}

\usepackage{graphicx}
\usepackage{dcolumn}

\usepackage{dsfont}
\usepackage[capitalise,sort,noabbrev]{cleveref}
\usepackage{csquotes}
\usepackage{graphicx}
\usepackage{color}
\usepackage{asypictureB}

\begin{asyheader}
	usepackage("amsmath");
\end{asyheader}

\usepackage{tikz}

\usetikzlibrary{positioning,arrows,decorations.markings,decorations.pathreplacing,shapes.geometric,fit,calc,shadings}
\usepackage[miktex]{gnuplottex}

\usepackage{pgfplots}
\pgfplotsset{compat=1.7}

\pgfplotsset{
	legend image with text/.style={
		legend image code/.code={%
			\node[anchor=center] at (0.3cm,0cm) {#1};
		}
	},
}

\newcommand{\bra}[1]{\left\langle #1 \right|}
\newcommand{\ket}[1]{\left| #1\right\rangle}

\newcommand{\abs}[1]{\vert #1 \vert}
\newcommand{\abssq}[1]{\vert #1 \vert^2}
\newcommand{\idone}{\hat{\mathds{1}}}

\newcommand{\vsk}{\vec{s}_{K}}
\newcommand{\vtk}{\vec{t}_{K}}

\newcommand{\pr}{\mathrm{pr}}

\renewcommand{\vec}[1]{\boldsymbol{#1}}

\DeclareMathOperator\diag{diag}

\begin{document}

	\title{The geometric link between Hardy and Clauser-Horne-Shimony-Holt}

	\author{Johannes Seiler}
	\affiliation{Institut f{\"u}r Quantenphysik \& Center for Integrated Quantum Science and Technology ($\mathrm{IQ^{ST}}$), Universit{\"a}t Ulm, D-89069 Ulm, Germany}
	\email{johannes.seiler@uni-ulm.de}
	\author{Thomas Strohm}%
	\affiliation{%
	Corporate Research, Robert Bosch GmbH, D-71272 Renningen, Germany	}%

	\author{Wolfgang P. Schleich}
	\affiliation{Institut f{\"u}r Quantenphysik \& Center for Integrated Quantum Science and Technology ($\mathrm{IQ^{ST}}$), Universit{\"a}t Ulm, D-89069 Ulm, Germany}%
	\affiliation{%
	Hagler Institute for Advanced Study, Institute for
	Quantum Science and Engineering (IQSE), and Texas A{\&}M AgriLife
	Research, Texas A{\&}M University, College Station, Texas 77843-4242, USA	}%
		
	\date{\today}
	
	\begin{abstract}
	We show that the Hardy nonlocality condition is equivalent to the violation of the CHSH inequality with additional constraints. We adapt the geometrical optimization of the violation of the CHSH inequality to these additional constraints and show that the Hardy condition is equivalent to optimizing the length difference of two sides in a triangle. Furthermore, we discuss the effects of the different constraints.
	\end{abstract}
	
	\maketitle
	
	
\section{Introduction}\label{sec:introduction}
A typical Bell test scenario \cite{Bell1964,Brunner2014,Freedman1972,Fry1976,Aspect1982a,Aspect1982,Weihs1998,Hensen2015,Giustina2015,Shalm2015} consists of two qubits and two independent projective measurements on each qubit. The measurements are chosen in such a way that the sum of expectation values of the four possible correlated measurements violate the Clauser-Horne-Shimony-Holt (CHSH) inequality \cite{Clauser1969}, which would hold classically.

A different approach to detect nonlocality was proposed by Hardy in 1993 \cite{Hardy1993}. 
In the Hardy scenario, the setup is similar, but four measurement outcomes are selected and three of their associated joint probabilities are fixed. Classically, the chosen probabilities lead to a vanishing outcome of the forth joint measurement. The nonlocality of quantum mechanics then manifests itself in a nonvanishing measurement outcome. 

In this article we establish a connection between the Hardy scenario and the CHSH inequality: The nonvanishing outcome in the Hardy scenario is proportional to the violation in the CHSH inequality under three additional constraints on the measurements. In order to demonstrate this notion, we first start from the Hardy scenario and show that one of the probabilities is given by the expectation value of the CHSH inequality, which when optimized with respect to three constraints yields the Hardy scenario. In this way we gain a geometrical insight of the Hardy scenario, which further allows us to find and comprehend the optimal measurement settings to maximize the probability of finding a violation of the classical result.

When we start from the CHSH inequality and apply more and more constraints to the measurement settings we again reach the Hardy case. This approach brings out most clearly the crucial role of the constraints in obtaining the Hardy nonlocality condition.

\subsection{Hardy scenario}
The idea proposed by Hardy is to construct a set of measurements on a bipartite state, such that a specific measurement outcome is predicted by quantum mechanics, which cannot be explained by classical physics. 

For this purpose, we consider a bipartite two qubit state. On the subsystem $ A $ we either perform the measurement $ Q $ or $ R $ which each can yield the outcome $ +1 $ or $ -1 $. Analogously, on the subsystem $ B $ we choose between the two measurements $ S $ and $ T $. 

The measurements are defined such that the joint probability measuring $ Q=1 $ and $ T=1 $ is zero. Furthermore, if one measures $ S=1 $, we always find $ Q=1 $ and analogously, if we measure $ R=1 $, one always finds $ T=1 $. The question is then, what is the joint probability to obtain $ R=1 $ and $ S=1 $ in a single measurement. 
\begin{figure}
	\centering
	\includegraphics{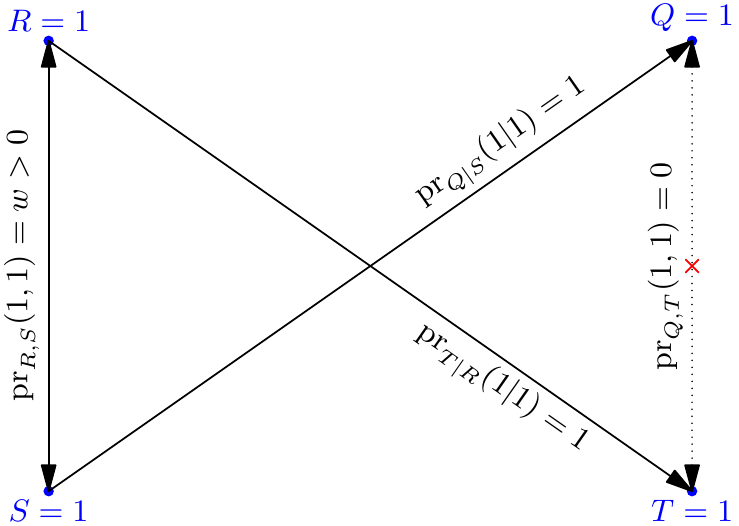}
	\caption{Schematic connection of the measurement results and contradition to classical local theory. The two points $ S=1 $ and $ R=1 $ can be measured in single measurement on both subsystems with probability $ w $. The conditional probabilities infer that one would also measure $ Q=1 $ and $ T=1 $ in a single run with the same probability. Instead this joint probability is zero and thus contradicts the nonlocality assumption. }
	\label{fig:probsquare}
\end{figure}

In any classical, local theory, the measurements on each subsystem do not influence each other and we change the measurement on a subsystem without changing the measurement outcome on the other subsystem. 

We now assume that we find an outcome $ R=1 $ and $ S=1 $ in a single measurement. Then changing the measurement on the subsystem $ A $ from $ R $ to $ Q $, we would obtain $ Q=1 $. In full analogy, if we change our measurement on the subsystem $ B $ from $ S $ to $ T $, we would measure $ T=1 $. 

As a consequence, if we changed the measurements on both subsystems simultaneously, the outcome of our new measurements would be $ Q=1 $ and $ T=1 $, which is in contradiction to the requirement, that the joint probability of $ Q=1 $ and $ T=1 $ is set to zero. Thus, for any classical theory, a common measurement of $ R=1 $ and $ S=1 $ cannot occur. This contradiction is schematically depicted in \cref{fig:probsquare}.

However, it can be shown that quantum theory allows to construct measurements such that they obey the all of three requirements mentioned above, and still have a nonvanishing probability $ w $ of finding $ R=1 $ and $ S=1 $ in the same run. We denote this situation as Hardy nonlocality. For a vivid interpretation of this scenario we refer to the work of Kwiat and Hardy \cite{Kwiat2000}.

Therefore, in contrast to Bell inequalities, such as the CHSH inequality, in principle only a single measurement can suffice to violate the classical predictions. This behavior is similar to the famous Greenberger-Horne-Zeilinger (GHZ) scenario \cite{Greenberger1990} for three qubits and its extensions to multipartite multidimensional systems \cite{Cerf2002}. However, in contrast to the GHZ model, the violation in the Hardy scenario only occurs probabilistically. Nevertheless, in any implementation, in order to guarantee the other three conditions, one theoretically needs to perform infinitely many measurements.

\subsection{Outline}
Our article is organized in the following way. In Sec. \ref{sec:Hardy-CHSH} we formulate the Hardy scenario in a quantum mechanical formalism and show how it is related to the CHSH inequality plus additional constraints. We then develop a geometrical interpretation of the probability $ w $ in Sec. \ref{sec:geometric}. We exploit this interpretation in Sec. \ref{sec:optimization} to optimize the nonlocality probability $ w $, and determine the optimal measurement settings in Sec. \ref{sec:optimization_meas}. In Sec. \ref{sec:comp-chsh}, we compare our findings for the Hardy scenario to the results for the CHSH scenario, and discuss how the different constraints in the Hardy scenario effect the results found. Finally, we provide a conclusion in \ref{sec:conclusion}. 

In order to keep our article self-containing but focused on the main results, we provide three appendices. In Appendix \ref{app:parallel} we calculate the relations between the different measurement vectors due to the additional constraints. We devote Appendix \ref{app:paraStheta} to obtaining an analytical expression for the expected correlation value in terms of spherical coordinates. We finally provide a derivation of the violation probability in terms of a length appearing in our geometrical picture in Appendix \ref{app:w-geo-tau}. 

\section{Connection between the Hardy scenario and the CHSH inequality} \label{sec:Hardy-CHSH}
In this section, we connect the Hardy scenario to the well-known Clauser-Horne-Shimony-Holt (CHSH) inequality. For this purpose, we first set up the quantum state and measurements. We then provide the quantum theoretical formulation of the CHSH inequality and the Hardy scenario. Thereafter, we show that these formulations are equivalent to one another with the exception that the Hardy scenario must obey three additional constraints on the measurements performed. We discuss the effects of these constraints on the measurement directions and the maximization of the nonlocality violation probability in the Hardy scenario. Finally, we numerically optimize this probability.

\subsection{Quantum state and measurements}
We first define the quantum mechanical state and measurements and translate the classical probabilities of the Hardy problem, presented in Section \ref{sec:introduction}, into quantum mechanical probabilities.

Throughout the remainder of this article, we consider a pure two-qubit state
\begin{equation}\label{eq:Psi-definition}
	\ket{\Psi} \equiv \sqrt{\frac{1+\sqrt{1-\mathcal{C}^{2}}}{2}}\ket{0}_{A}\ket{0}_{B}-\sqrt{\frac{1-\sqrt{1-\mathcal{C}^{2}}}{2}}\ket{1}_{A}\ket{1}_{B}
\end{equation}
in the Schmidt basis \cite{JanosA.Bergou2013}, where $ \ket{0} $ and $ \ket{1} $ are the eigenstates of the Pauli $ \hat{\sigma}_{z} $ operator with eigenvalue $ +1 $ and $ -1 $ in the respective subsystem. Furthermore, the concurrence $ \mathcal{C} $ \cite{Wootters1998,Rungta2001}, with $ 0\leq \mathcal{C} \leq 1 $ is a measure of the entanglement between the two subsystems of the two-qubit state. For $ \mathcal{C}=0 $ the state is separable, while $ \mathcal{C}=1 $ is a maximally entangled Bell state. 

We note that the state $ \ket{\Psi} $ is an arbitrary state with given entanglement, that is any other state with the same entanglement can be transformed into \cref{eq:Psi-definition} by a change of basis.

Due to the symmetry of the state $ \ket{\Psi} $ in the Schmidt basis, the Bloch vectors $ \vec{a}_{A} = \bra{\Psi} \hat{\vec{\sigma}}_{A}\ket{\Psi}$ and $ \vec{a}_{B}= \bra{\Psi} \hat{\vec{\sigma}}_{B}\ket{\Psi} $ of the two subsystems $ A $ and $ B $, where $ \hat{\vec{\sigma}} $ is the vector of the Pauli matrices $ \hat{\sigma}_{x} $, $ \hat{\sigma}_{y} $ and $ \hat{\sigma}_{z} $, have the same representation
\begin{equation}\label{eq:veca}
	\vec{a}_{A}=\vec{a}_{B} = \begin{pmatrix}
		0\\ 0\\ \sqrt{1-\mathcal{C}^{2}}
	\end{pmatrix}\equiv\vec{a}
\end{equation}
and we drop the indices in the following.

On each of the two subsystems we perform one of two possible measurements. On the subsystem $ A $, we choose between the operators
\begin{equation}\label{eq:QandR-definition}
	\hat{Q} = \vec{q}\cdot\hat{\vec{\sigma}}_{A} \quad \mathrm{or} \quad \hat{R} = \vec{r}\cdot\hat{\vec{\sigma}}_{A},
\end{equation}
where $ \vec{q} $ and $ \vec{r} $ are three-dimensional unit vectors that denote the measurement direction on the Bloch sphere of the subsystem $ A $.

In full analogy, on the subsystem $ B $ we perform either the measurement 
\begin{equation}\label{eq:SandT-definition}
	\hat{S} = \vec{s}\cdot\hat{\vec{\sigma}}_{B} \quad \mathrm{or} \quad \hat{T} = \vec{t}\cdot\hat{\vec{\sigma}}_{B},
\end{equation}
with the three-dimensional unit vectors $ \vec{s} $ and $ \vec{t} $. 

\subsection{CHSH inequality}
From these measurement settings, we derive the sum of expectation values  
\begin{equation}\label{eq:S-avg-Operators}
	\mathcal{S} \equiv \langle \hat{Q}\otimes \hat{S} \rangle- \langle \hat{Q}\otimes \hat{T} \rangle + \langle \hat{R}\otimes \hat{S} \rangle + \langle \hat{R}\otimes \hat{T} \rangle
\end{equation}
of the individual measurement correlations. Inserting the state $ \ket{\Psi} $, \cref{eq:Psi-definition}, and the measurement operators, \cref{eq:QandR-definition,eq:SandT-definition}, into \cref{eq:S-avg-Operators}, we obtain the expectation value
\begin{equation}\label{eq:S-avg-vectors}
	\mathcal{S} = \vec{q}^{T}K(\vec{s}-\vec{t}) + \vec{r}^{T}K(\vec{s}+\vec{t})
\end{equation}
in terms of the measurement vectors $ \vec{q} $, $\vec{r} $, $\vec{s} $ and $ \vec{t} $ and the correlation matrix
\begin{equation}\label{key}
	K \equiv \bra{\Psi}\hat{\vec{\sigma}}_{A}\otimes \hat{\vec{\sigma}}_{B} \ket{\Psi},
\end{equation}
which is a property of the state alone.
For the state $ \ket{\Psi} $, \cref{eq:Psi-definition}, the correlation matrix is explicitly given by 
\begin{equation}\label{eq:K-diag}
	K = \diag(-\mathcal{C},\mathcal{C},1),
\end{equation}
where the diagonal form of the matrix is a direct consequence of using the Schmidt decomposition. Note, that $ K\vec{a} = \vec{a}$, since $ \vec{a} $ points along the $ z $ axis.

For any classically correlated, local system, the CHSH inequality
\begin{equation}\label{eq:CHSH-inequality}
	\mathcal{S}\leq 2
\end{equation}
holds. However, \cref{eq:CHSH-inequality} can be violated for entangled quantum mechanical states. Indeed, for a state with concurrence $ \mathcal{C} $, the expectation value is bounded by
\begin{equation}\label{eq:Smax-CHSH}
	\mathcal{S}_{\mathrm{CHSH}} \leq 2\sqrt{1+\mathcal{C}^{2}}
\end{equation}
for any CHSH like experiment. Only for separable states ($ \mathcal{C}=0 $) a violation of the CHSH inequality is impossible. For maximally entangled states, that is $ \mathcal{C}=1 $, we find $ \mathcal{S} \leq 2\sqrt{2} $, the well-known Tsirelson bound \cite{Cirelson1980,Tsirelson1987}. 
%

\subsection{Quantum probabilities in the Hardy setting}
In contrast to the CHSH inequality, the idea of the Hardy scenario is to prepare the above measurements in such a way, that we guarantee them to fulfill the three probability distributions for the outcomes $ Q,R,S $ and $ T $ associated with a single measurement of the operator $ \hat{Q},\hat{R},\hat{S} $ and $ \hat{T} $, discussed in Section \ref{sec:introduction}.

The first condition is that the joint probability
\begin{equation}\label{eq:prob-hardy-1}
	\pr_{Q,T}(1,1)\equiv \bra{\Psi}\frac{1}{2}(\idone+\hat{Q})\otimes\frac{1}{2}(\idone+\hat{T})\ket{\Psi}  = 0,
\end{equation}
of simultaneous measurement of $ \hat{Q} $ and $ \hat{T} $ on the two subsystems of the state $ \ket{\Psi} $ cannot yield the outcome $ 1 $ on both sides. 

The second requirement is finding the outcome $ S=1 $ when measuring $ \hat{S} $ on subsystem $ B $ will definitely result in the outcome $ Q=1 $ for a measurement of $ \hat{Q} $, that is we have the conditional probability
\begin{equation}\label{eq:prob-hardy-2}
	\pr_{Q|S}(1|1)\equiv \frac{\bra{\Psi}\frac{1}{2}(\idone+\hat{Q})\otimes\frac{1}{2}(\idone+\hat{S})\ket{\Psi}}{\bra{\Psi}\frac{1}{2}(\idone+\hat{S})\ket{\Psi}} = 1,
\end{equation}

In full analogy to the second condition, the third is the conditional probability
\begin{equation}\label{eq:prob-hardy-3}
	\pr_{T|R}(1|1)\equiv \frac{\bra{\Psi}\frac{1}{2}(\idone+\hat{R})\otimes\frac{1}{2}(\idone+\hat{T})\ket{\Psi}}{\bra{\Psi}\frac{1}{2}(\idone+\hat{R})\ket{\Psi}} =1.
\end{equation}
which predicts that we find $ T=1 $ on the subsystem $ B $, when we measured $ R=1 $ on the subsystem $ A $. 

We then ask the question, what is the probability
\begin{equation}\label{eq:prob-hardy-4}
	w\equiv \pr_{R,S}(1,1) \equiv \bra{\Psi}\frac{1}{2}(\idone+\hat{R})\otimes\frac{1}{2}(\idone+\hat{S})\ket{\Psi}
\end{equation}
of the joint measurement outcomes $ R=1 $ and $ S=1 $. 


As we have discussed in the introduction, if our measurement settings fulfill \cref{eq:prob-hardy-1,eq:prob-hardy-2,eq:prob-hardy-3}, obtaining a single measurement outcome with $ R=1 $ and $ S=1 $ in the same measurement is a violation of any local hidden variable theory.

Inserting the definitions of our state $ \ket{\Psi} $, \cref{eq:Psi-definition}, and measurement operators, \cref{eq:QandR-definition,eq:SandT-definition}, into the definition of the joint and conditional probabilities, \cref{eq:prob-hardy-1,eq:prob-hardy-2,eq:prob-hardy-3,eq:prob-hardy-4}, we obtain a set of four equations:
\begin{align}
	\vec{q}^{T}K\vec{t} &= -\left(1+\vec{a}\cdot \vec{q} + \vec{a}\cdot \vec{t}\right) \label{eq:prob-vectors-1}, \\ \vec{q}^{T}K\vec{s}&= 1-\vec{a}\cdot \vec{q}+\vec{a}\cdot \vec{s} \label{eq:prob-vectors-2},\\
	\vec{r}^{T}K\vec{t}&=1+\vec{a}\cdot \vec{r}-\vec{a}\cdot \vec{t} ,\label{eq:prob-vectors-3} \\
	\vec{r}^{T}K\vec{s}&= -\left(1+\vec{a}\cdot \vec{r} + \vec{a}\cdot \vec{s}\right)+ 4w, \label{eq:prob-vectors-4}
\end{align}
which only depend on the measurement vectors $ \vec{q},\vec{r},\vec{s} $ and $ \vec{t} $, the Bloch vector $ \vec{a} $, and the correlation matrix $ K $.

\subsection{Relation to the CHSH inquality}
We are now in the position to compare the Hardy scenario with the CHSH inequality. In order to do so, we add \cref{eq:prob-vectors-2,eq:prob-vectors-3,eq:prob-vectors-4} and subtract \cref{eq:prob-vectors-1}, and obtain
\begin{equation}\label{eq:sum-probs}
	 \vec{q}^{T}K(\vec{s}-\vec{t})+\vec{r}^{T}K(\vec{s}+\vec{t}) = 4w - 2 .
\end{equation}
We identify the left hand side of \cref{eq:sum-probs} as the expectation value $ \mathcal{S} $, \cref{eq:S-avg-vectors}, at the heart of the CHSH inequality. 

Hence, the probability
\begin{equation}\label{eq:w-ito-S}
	w = \frac{\mathcal{S}-2}{4}
\end{equation}
of finding a measurement outcome that contradicts the locality condition is directly connected to the violation of the CHSH inequality, \cref{eq:CHSH-inequality}.  If and only if the measurement setting would violate this inequality, it is possible to obtain a measurement of $ Q=1 $ and $ S=1 $ in the same run. For this reason, we call $ w $ the violation probability.

In order to maximize $ w $, we simply have to maximize the expectation value $ \mathcal{S} $, which corresponds to maximizing the violation of the CHSH inequality.

The maximization of $ \mathcal{S} $ over the four independent measurements $ \vec{q} $, $ \vec{r} $, $ \vec{s} $ and $ \vec{t} $ for a pure state is known \cite{Gisin1991}. We even demonstrated a geometrical approach \cite{Seiler2021} which allows one to find all possible measurement that maximize $ \mathcal{S} $ for a given concurrence $ \mathcal{C} $. 

However, in addition to \cref{eq:w-ito-S}, the probabilities \cref{eq:prob-hardy-1,eq:prob-hardy-2,eq:prob-hardy-3} between correlated measurement outcomes still must be fulfilled. As a consequence the measurement vectors $ \vec{q} $, $ \vec{r} $, $ \vec{s} $ and $ \vec{t} $ are no longer independent of each other. 

Unfortunately, as it turns out none of the infinitely many optimal measurement settings for maximizing $ \mathcal{S} $ for a given concurrence $ \mathcal{C} $ without any further constraints fulfills the three additional constraints, \cref{eq:connection-qt,eq:connection-qs,eq:connection-rt}, on the measurement vectors. In order to maximize $ w $, the expectation value thus has to be maximized again under these new constraints.

\subsection{Constraints for measurement directions}
We now take a closer look at the additional constraints, which arise from demanding the probabilities \cref{eq:prob-hardy-1,eq:prob-hardy-2,eq:prob-hardy-3}. 

The first constraint, \cref{eq:prob-vectors-1}, leads to 
\begin{equation}\label{key}
	(K\vec{q}+\vec{a})\cdot \vec{t} = -(1+\vec{a}\cdot \vec{q}),
\end{equation}
which we rewrite as
\begin{equation}\label{key}
	\abs{K\vec{q}+\vec{a}}\abs{\vec{t}} \cos\alpha = - \abs{1+\vec{a}\cdot \vec{q}},
\end{equation}
where $ \alpha $ is the angle between $ \vec{t} $ and $ K\vec{q}+\vec{a} $. 
The premise of $ \vec{t} $ being a unit vector, together with the relation 
\begin{equation}\label{key}
	\abs{K\vec{q}+\vec{a}} = \abs{1+\vec{a}\cdot \vec{q}},
\end{equation}
which is proven in Appendix \ref{app:parallel}, then constitute that the measurement vector $ \vec{t} $ is antiparallel to $ K\vec{q}+\vec{a} $, leading to the connection
\begin{equation}\label{eq:connection-qt}
	\vec{t} = - \frac{K\vec{q}+\vec{a}}{\abs{K\vec{q}+\vec{a}}}
\end{equation}
between the vectors $ \vec{q} $ and $ \vec{t} $. 

The second constraint, \cref{eq:prob-vectors-2}, leads to the connection
\begin{equation}\label{eq:connection-qs}
	\vec{s} =  \frac{K\vec{q}-\vec{a}}{\abs{K\vec{q} -\vec{a}}}
\end{equation}
between the vectors $ \vec{q} $ and $ \vec{s} $. 

Analogously, we find 
\begin{equation}\label{eq:connection-rt}
	\vec{r} =  \frac{K\vec{t}-\vec{a}}{\abs{K\vec{t} -\vec{a}}},
\end{equation}
between the vectors $ \vec{r} $ and $ \vec{t} $ from the third constraint, \cref{eq:prob-vectors-3}. 

As a consequence, instead of four independent vectors $ \vec{q}$, $\vec{r}$, $\vec{s} $ and $ \vec{t} $, the three vectors $\vec{r}$, $\vec{s} $ and $ \vec{t} $ are fully determined by the vector $ \vec{q} $. We only have one independent vector  $ \vec{q} $ left. 

\subsection{Numerical maximization of the probability $ w $}
The expectation value $ \mathcal{S} $, \cref{eq:S-avg-vectors},  under the constraints \cref{eq:connection-qt,eq:connection-qs,eq:connection-rt}, and hence the probability $ w $, \cref{eq:w-ito-S}, is a function of only a single measurement vector. 

In order to maximize the probability $ w $, we parameterize the measurement vector
\begin{equation}\label{eq:q-spherical-coordinates}
	  \vec{q} =\begin{pmatrix}
	  	\cos\phi \sin\theta \\ \sin\phi \sin\theta \\ \cos \theta
	  \end{pmatrix} 
\end{equation}
in spherical coordinates by two angles $ \theta $ and $ \phi $ and insert the resulting expression together with the expressions for the other three vectors $ \vec{r},\vec{s},\vec{t} $, \cref{eq:connection-qs,eq:connection-qt,eq:connection-rt}, and the correlation matrix $ K $, \cref{eq:K-diag}, into the definition of the expectation value $ \mathcal{S} $, \cref{eq:S-vectors-K}. The resulting expression for the expectation value, given in Appendix \ref{app:paraStheta}, only depends on the angle $ \theta $ and the concurrence $ \mathcal{C} $, but is rather cumbersome and difficult to maximize analytically.

\begin{figure}
	\centering
	\includegraphics{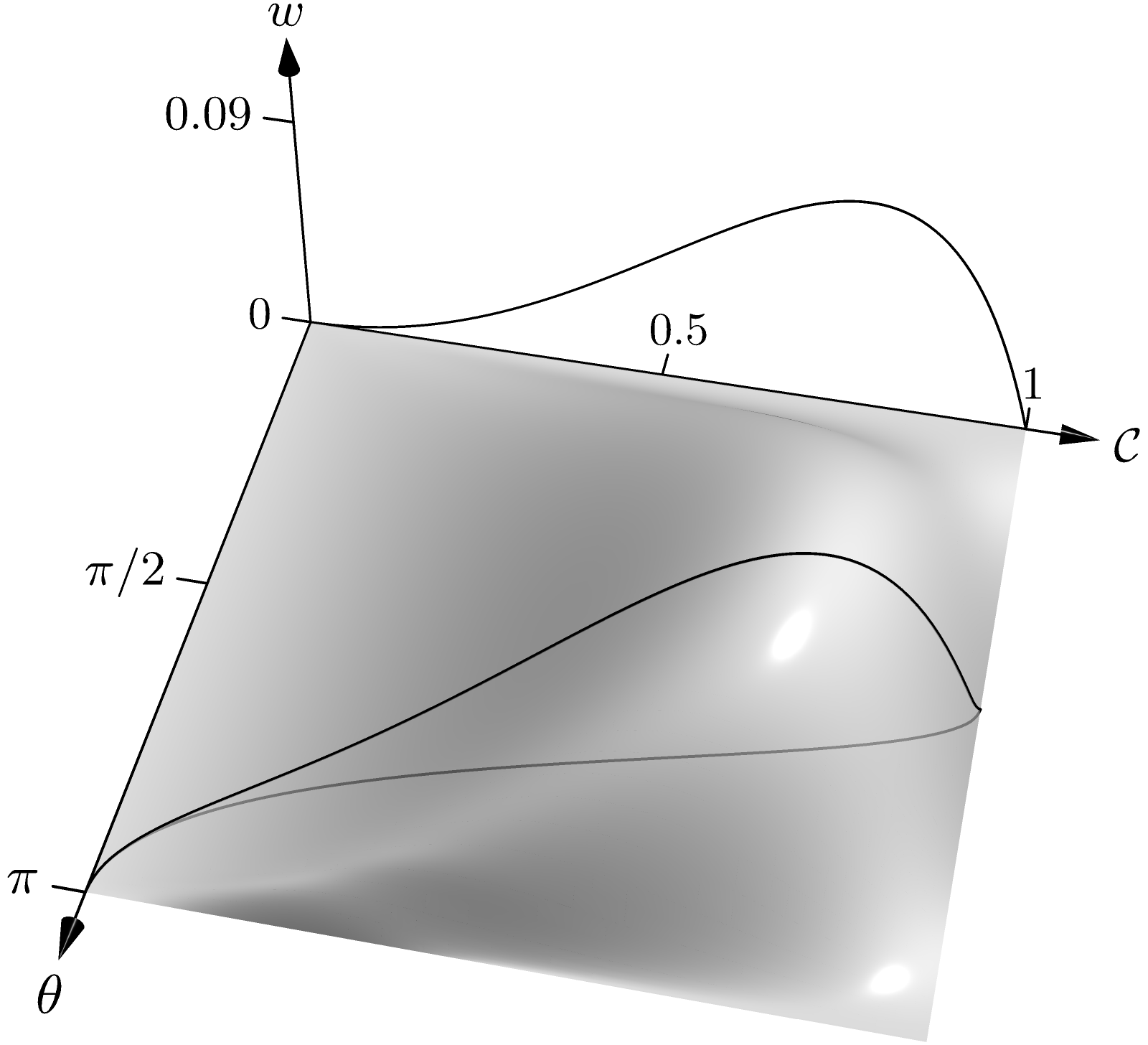}
	\caption{The probability $ w $ as a function of the polar angle $ \theta $ of the vector $ \vec{q} $ and the concurrence $ \mathcal{C} $. For both separable ($ \mathcal{C}=0 $) and maximally entangled states ($ \mathcal{C}=1 $), as well as for $ \vec{q} $ pointing along the positive ($ \theta= 0$) or negative ($ \theta= \pi/2$) $ z $ axis, the probability $ w $ vanishes. The black line marks the optimal $ w $ as a function of the concurrence $ \mathcal{C} $. The grey line indicates the projection of the optimal line in the $ \mathcal{C} $-$\theta$ plane. }
	\label{fig:w-func}
\end{figure}
Therefore, we first analyze this result numerically. \Cref{fig:w-func} shows the probability $ w $ as a function of the polar angle $ \theta $ and the concurrence $ \mathcal{C} $. 

As a result, we obtain that the violation probability $ w $ is vanishing for either separable states, that is $ \mathcal{C}=0 $, or maximally entangled states ($ \mathcal{C}=1 $). While the vanishing probability for $ \mathcal{C}=0 $ is expected, since entanglement is a key ingredient in the violation of the CHSH inequality, the case of a maximally entangled state is more of a surprise, as the CHSH inequality is in general maximally violated for a maximally entangled state. Hence, the vanishing probability must occur due to the additional constraints on the measurement vectors.

Moreover, the probability $ w $ also vanishes if the polar angle of the measurement vector $ \vec{q} $ is either $ \theta = 0 $ or $ \theta=\pi/2 $. For these cases the vector $ \vec{q} $ points along the positive or negative $ z $ axis, respectively. This result is also not true for the general CHSH case, and thus must result from the additional constraints.

Furthermore, for any fixed concurrence $ 0 < \mathcal{C} < 1 $ a unique angle $ \theta $ exists, such that the probability $ w $ is maximal. These values are indicated by the black line in \cref{fig:w-func}. For small concurrences the optimal angle is close to $ \pi $, that is the optimal vector $ \vec{q} $ points mostly along the negative $ z $ axis. With increasing $ \mathcal{C} $ the angle $ \theta $ decreases monotonically towards $ \pi/2 $, that is the vector $ \vec{q} $ moves towards the $ x $-$ y $ plane. For all concurrences the optimal vector $ \vec{q} $ thus has a negative $ z $ component. 



\section{Geometric interpretation}\label{sec:geometric}
In the previous section we have shown that the violation probability $ w $ in the Hardy scenario is related to the expectation value $ \mathcal{S} $ of the CHSH inequality with three additional constraints on the measurement vectors. The optimization of the probability is thus a optimization of the CHSH setup under constraints.

Moreover, we have performed a numerical evaluation of the probability $ w $, in which we have seen that the probability is non-vanishing as long as the state is neither unentangled nor maximally entangled, that is $ 0 < \mathcal{C} < 1 $, and the vector $ \vec{q} $ does not point along neither the $ z $ axis nor in the $ x $-$ y $ plane. Furthermore, there exists a unique maximum of the probability for any given concurrence $ \mathcal{C} $, for which the vector $ \vec{q} $ always has a negative $ z $ component.

In Ref. \cite{Seiler2021}, we have demonstrated that the optimal expectation value $ \mathcal{S} $ has a geometrical interpretation as half of the perimeter of a parallelogram enclosed by an ellipse whose semimajor and semiminor axes have lengths $ 1 $ and $ \mathcal{C} $, respectively. This interpretation allows us to efficiently and analytically find all possible optimization strategies for the expectation value $ \mathcal{S} $. 

In this section we adapt our geometric picture to the Hardy scenario, and apply it to explain the results found in our numerical analysis in the previous section. In particular we first show that all the relevant measurement vectors lie in a common plane. This behavior explains the independence of the probability $ w $ on the azimuthal angle $ \phi $ of the vector $ \vec{q} $. We then show that for all instances in which the probability $ w $ vanishes, the plane collapses to a single line. Moreover, we demonstrate how the relevant vectors are constructed geometrically. Finally, we exploit this construction to provide a geometrical interpretation of the probability $ w $ as the length difference in a triangle, and use this description to parameterize the probability. 

\subsection{Measurement vectors lie in a plane}
In our numerical calculation of the probability $ w $, we have seen that it is independent of the azimuthal angle $ \phi $ of the vector $ \vec{q} $. The probability is therefore symmetric under rotations around the $ z $ axis. In this section we show this symmetry by proofing that all relevant vectors lie in a common plane, which contains the $ z $ axis of our coordinate system.

In the expectation value $ \mathcal{S} $, \cref{eq:S-avg-vectors}, only combinations of a vector of the subsystem $ A $ and a vector of the subsystem $ B $ appear, while the correlation matrix $ K $ specifies how vectors from different subsystems are multiplied to each other. 

Instead of using the correlation matrix as a connection between the two subsystems, we directly apply the correlation matrix to the vectors of the subsystem $ B $, defining the new vectors
\begin{equation}\label{eq:vsk-def}
	\vsk \equiv K\vec{s}
\end{equation}
and 
\begin{equation}\label{eq:vtk-def}
	\vtk \equiv K\vec{t}.
\end{equation}
These new vectors are no longer of unit length. Instead, their length depends on their orientation, since their components in the $ x $ and $ y $ direction are contracted by the factor $ \mathcal{C} $.

The new vectors $ \vsk $ and $ \vtk $ directly multiply to the vectors $ \vec{q} $ and $ \vec{r} $ of the subsystem $ A $ by the usual scalar product, that is we find
\begin{equation}\label{eq:S-vectors-K}
	\mathcal{S} = \vec{q}\cdot(\vsk -\vtk) + \vec{r}\cdot(\vsk+\vtk)
\end{equation}
for the expectation value.

Our aim is to construct the vectors $ \vec{q},\vec{r},\vsk $ and $ \vtk $ geometrically. For this purpose we first show that these vectors lie in a common plane, defined by the vector 
\begin{equation}\label{key}
	 K^{2}\vec{q} = \begin{pmatrix}
	 	\mathcal{C}^{2}q_{x}\\ \mathcal{C}^{2}q_{y} \\ q_{z}
	 \end{pmatrix}, 
\end{equation} 
which is in general not parallel to the vector $ \vec{q} $, and the Bloch vector $ \vec{a} $. 

From the definitions of $ \vsk $ and $ \vtk $, \cref{eq:vsk-def,eq:vtk-def}, and of $ \vec{s} $ and $ \vec{t} $, \cref{eq:connection-qs,eq:connection-qt}, the new vectors $ \vsk $ and $ \vtk $ are given by the explicit expressions
\begin{equation}\label{eq:vsk-explicit}
	\vsk = \frac{K^{2}\vec{q}-\vec{a}}{\abs{K\vec{q}-\vec{a}}}
\end{equation}
and
\begin{equation}\label{eq:vtk-explicit}
	\vtk = -\frac{K^{2}\vec{q}+\vec{a}}{\abs{K\vec{q}+\vec{a}}}
\end{equation}
where we made use of the fact that the correlation matrix $ K $ does not change the Bloch vector $ \vec{a} $, since the latter points along the $ z $ axis of our coordinate system.

In general the Bloch vector $ \vec{a} $ is fixed, while the measurement vector $ \vec{q} $ is our free parameter. For any arbitrary but fixed $ \vec{q} $, it immediately follows form \cref{eq:vsk-explicit,eq:vtk-explicit}, that $ \vsk $ and $ \vtk $ are in the plane $ P $ spanned by the vectors $ K^{2}\vec{q} $ and $ \vec{a} $.

Furthermore, from \cref{eq:connection-rt,eq:vtk-def} it follows that the vector
\begin{equation}\label{key}
	\vec{r} = \frac{\vtk-\vec{a}}{\abs{\vtk-\vec{a}}}
\end{equation}
is a linear combination of the two vectors $ \vtk $ and $ \vec{a} $. Therefore, the vector $ \vec{r} $ also lies in the plane $ P $.
%

Moreover, since the measurement vector $ \vec{q} $ can be rewritten as  
\begin{equation}\label{key}
	 \vec{q} = \frac{K^{2}\vec{q}-(\vec{q}\cdot\vec{a})\vec{a}}{\mathcal{C}^{2}},
\end{equation} 
which directly follows form evaluating the right hand side, $ \vec{q} $ is also a linear combination of $ \vec{a} $ and $ K^{2}\vec{q} $. The plane $ P $ thus contains the vector $ \vec{q} $. 

Therefore all the four vectors appearing in \cref{eq:S-vectors-K} lie in $ P $ and we thus restrict our discussions and calculations for the remainder of our article to the plane $ P $. 

We finally note that the Bloch vector $ \vec{a} $ is parallel to the $ z $ axis of our coordinate system, and the plane $ P $ thus contains the $ z $ axis. Therefore, the second direction orthogonal on $ z $ is a vector in the $ x $-$ y $ plane, which we denote by $ x^{\prime} $. Hence, $ P $ is the $ x^{\prime} $-$ z $ plane. From this form of $ P $ it directly follows that the expectation value  $ \mathcal{S} $, \cref{eq:S-vectors-K}, is independent of the azimuthal angle $ \phi $ of the vector $ \vec{q} $, which is consistent with our results from the previous section.

\subsection{Degenerate plane leads to vanishing violation probability}
The plane $ P $ is only uniquely defined as long as the vectors $ K^{2}\vec{q} $ and $ \vec{a} $ are neither parallel nor either of them vanishes. We now examine the cases where this condition is not fulfilled.


The first case we consider is when $ \vec{a} $ is vanishing. According to its definition, \cref{eq:veca}, the Bloch vector only vanishes for $ \mathcal{C}=1 $, which furthermore implies $ K = \diag(-1,1,1) $. Inserting these conditions into \cref{eq:connection-qs,eq:connection-qt,eq:connection-rt}, we obtain $ \vec{q}=\vsk = -\vtk = -\vec{r} $, and thus all relevant vectors point along the same line. From our expression of the expectation value, \cref{eq:S-vectors-K}, it immediately follows under these conditions that $ \mathcal{S}=2 $, and therefore $ w = 0 $.

The second case we consider is that both $ K^{2}\vec{q} $ and $ \vec{a} $ are parallel, which corresponds to the vector $ \vec{q} $ pointing along the $ z $ axis of our coordinate system. 
It immediately follows from \cref{eq:connection-qs,eq:connection-qt,eq:connection-rt}, that the three other vectors $ \vec{r}, \vsk $ and $ \vtk $ also point along the $ z $-axis of our coordinate system. By the same argument as for the case $ \mathcal{C}=1 $, it immediately follows, that $ w=0 $. 

Finally, we consider the case of $ K^{2}\vec{q} = 0$, which is only achievable for $ \mathcal{C}=0 $. In this case the underlying state $ \ket{\Psi} $ is separable. Since a separable state never violates the CHSH inequality, the violation probability $ w $ is also vanishing.

Hence, the vectors $ \vec{q}, \vec{r}, \vsk $ and $ \vtk $ are either all located in a common well defined plane $ P $ or they point along a single line. In the latter case, the probability $ w $ always vanishes. Since we are only interested in obtaining a nonvanishing probability $ w $, we restrict ourselves in the following to the case of all four vectors $ \vec{q}, \vec{r}, \vsk $ and $ \vtk $ lying in a common plane $ P $.



\subsection{Geometric construction of the measurement vectors}
\begin{figure}
	\centering
	\includegraphics{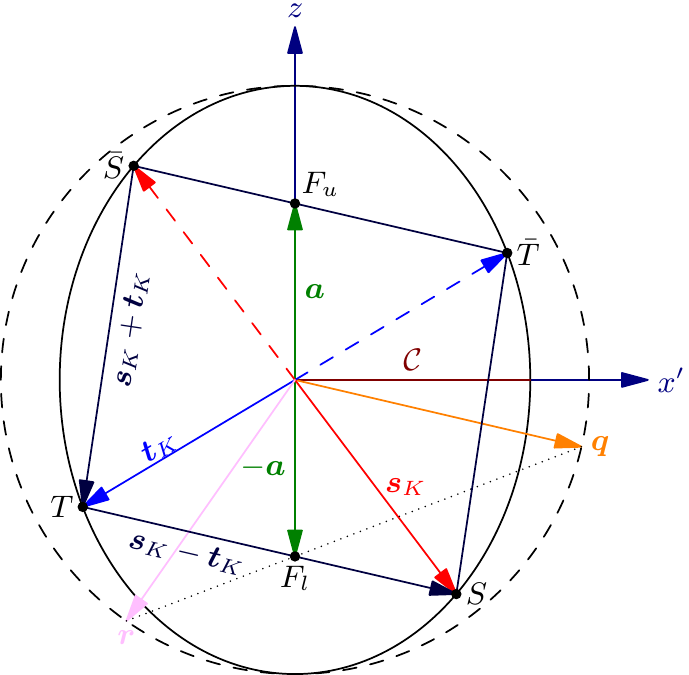}
	\caption{Geometric construction of the four vectors $ \vec{q} $, $ \vec{r} $, $ \vsk $ and $ \vtk $. The vector $ \vec{q} $ is a unit vector and points from the origin to a unit circle. The vectors $ \vsk $ and $ \vtk $ are no longer unit vectors but point from the origin to the points $ S $ and $ T $ on an ellipse with semi-major axis of unit length along the $ z $-axis and semi-minor axis of length $ \mathcal{C} $ along the $ x^{\prime} $-axis. The Bloch vectors $ \vec{a} $ and $ -\vec{a} $ point from the origin to the two focal points $ F_{u} $ and $ F_{l} $ of the ellipse. The points $ S $ and $ T $ are constructed by drawing a line parallel to the vector $ \vec{q} $ through the lower focal point $ F_{l} $ of the ellipse. The two intersections of this line with the ellipse are the points $ S $ and $ T $. Similarly, we can construct the points $ \bar{S} $ and $ \bar{T} $ by drawing a line parallel to $ \vec{q} $ through the upper focal point $ F_{u} $. The vector $ \vec{r} $ is constructed by drawing a line through the tip of the vector $ \vec{q} $ and the focal point $ F_{l} $ (dotted black line). The intersection with the unit circle constitutes the tip of the vector $ \vec{r} $. }
	\label{fig:geo-int}
\end{figure}

We are now in the position to construct the four vectors $ \vec{q} $, $ \vec{r} $, $ \vsk $ and $ \vtk $ in the plane $ P $ geometrically. This geometric interpretation is depicted in \cref{fig:geo-int}.

We start with the vector $ \vec{q} $, which is the only vector that can be chosen freely. Since $ \vec{q} $ is a unit vector, it points from the origin of our new coordinate system to a unit circle. 

The next vector we consider is the vector $ \vsk $, which is given by \cref{eq:vsk-explicit}. When starting at the origin the vector $ \vsk $ terminates in a point $ S $ on an ellipse with semi major axis of unit length along the $ z $ axis, and semi minor axis of length $ \mathcal{C} $. Using the representation of the Bloch vector $ \vec{a} $, \cref{eq:veca}, the correlation matrix $ K $, \cref{eq:K-diag}, and the fact that $ \vec{q} $ is a unit vector, we show in Appendix \ref{app:parallel} that the vector 
\begin{equation}\label{eq:sk-xi}
	\vsk = \xi \vec{q} - \vec{a},
\end{equation}
defines a line spanned by the support vector $ -\vec{a} $ and the directional vector $ \vec{q} $. Here, we introduced the abbreviation 
\begin{equation}\label{eq:def-xi}
	\xi \equiv \frac{\mathcal{C}^{2}}{1-\sqrt{1-\mathcal{C}^{2}}q_{z}} = \abs{\vsk+\vec{a}} 
\end{equation}
with $ q_{z} $ being the component of  $\vec{q}$ along the $ z $ axis.

We therefore geometrically construct the vector $ \vsk $, and thus the point $ S $, by going from the lower focal point $ F_{l} $ in the direction of $ \vec{q} $. The intersection with the ellipse constitutes the point $ S $.

In an analogous way, we determine the point $ T $ on the ellipse, which is defined by the vector $ \vtk $ pointing from the origin to the ellipse. By using the definition of $ \vec{t} $, \cref{eq:connection-qt}, as well as the Bloch vector $ \vec{a} $, and the correlation matrix $ K $, \cref{eq:K-diag}, we demonstrate in Appendix \ref{app:parallel} that the vector
\begin{equation}\label{eq:def-tau}
	\vtk =  -\tau \vec{q} - \vec{a},
\end{equation}
where we introduced the abbreviation 
\begin{equation}\label{eq:def-tau-ex}
	 \tau \equiv \frac{\mathcal{C}^{2}}{1+\sqrt{1-\mathcal{C}^{2}}q_{z}} = \abs{\vtk+\vec{a}} ,
\end{equation} 
is constructed by going from the lower focal point $ F_{l} $ in the direction $ -\vec{q} $ to the ellipse.

Our construction directly shows that, as a consequence of the additional constraints, \cref{eq:connection-qs,eq:connection-qt}, the line $ ST $, which also defines the vector $ \vsk-\vtk $, intersects the $ z $ axis in the focal point $ F_{l} $ of the ellipse.

From the vectors $ \vsk $ and $ \vtk $ we construct the vectors $ -\vsk $ and $ -\vtk $, which point from the origin to the points $ \bar{S} $ and $ \bar{T} $, respectively, on the ellipse. Due to symmetry, the line $ \bar{S}\bar{T} $, which also determines to the vector $ \vsk -\vtk $, intersects the $ z $ axis in the upper focal point $ F_{u} $.

The only vector left is the vector $ \vec{r} $. 
In Appendix \ref{app:parallel} we derive that this vector is given by the linear combination
\begin{equation}\label{eq:q-r}
	\vec{r} = \vec{q} - \frac{2}{2-\tau}\left(\vec{q}+\vec{a}\right)
\end{equation}
of the measurement vector $ \vec{q} $ and the Bloch vector $ \vec{a} $.
This expression shows that the vector $ \vec{r} $ can be constructed by going from the origin of our coordinate system to the end point of the vector $ \vec{q} $ and then along the direction $ -(\vec{q}+\vec{a}) $, which is through the focal point $ F_{l} $, until one again intersects the unit circle. This intersection is then the end point of the vector $ \vec{r} $. As a consequence of this construction, it is obvious that the line between the two end points of the vectors $ \vec{q} $ and $ \vec{r} $ on the unit circle includes the focal point $ F_{l} $.

This behaviour is similar to the construction of the points $ S $ and $ T $, where the line $ ST $ also includes the same focal point $ F_{l} $. 

We furthermore notice, that the vector $ \vec{r} $ is by its definition, \cref{eq:connection-rt}, parallel to the vector $ \vtk-\vec{a} $, and therefore the line $ \bar{T}F_{l} $ between the point $ \bar{T} $, defined by the vector $ -\vtk $, on the ellipse and the focal point $ F_{l} $. 

\subsection{Violation probability is the length difference of two sides in a triangle}
We now utilize the geometrical picture of the vectors $ \vec{q} $, $ \vec{r} $, $ \vsk $ and $ \vtk $ to find a geometrical interpretation of the violation probability $ w $. We start by determining the expectation value $ \mathcal{S} $ in our geometrical picture. We then use this result to relate the probability $ w $ to the length difference in a triangle.

The expectation value $ \mathcal{S} $, \cref{eq:S-avg-vectors}, consist of the two terms $ \vec{q}\cdot (\vsk-\vtk) $ and $ \vec{r}\cdot (\vsk+\vtk) $. 
%
\begin{figure}
	\centering
	\includegraphics{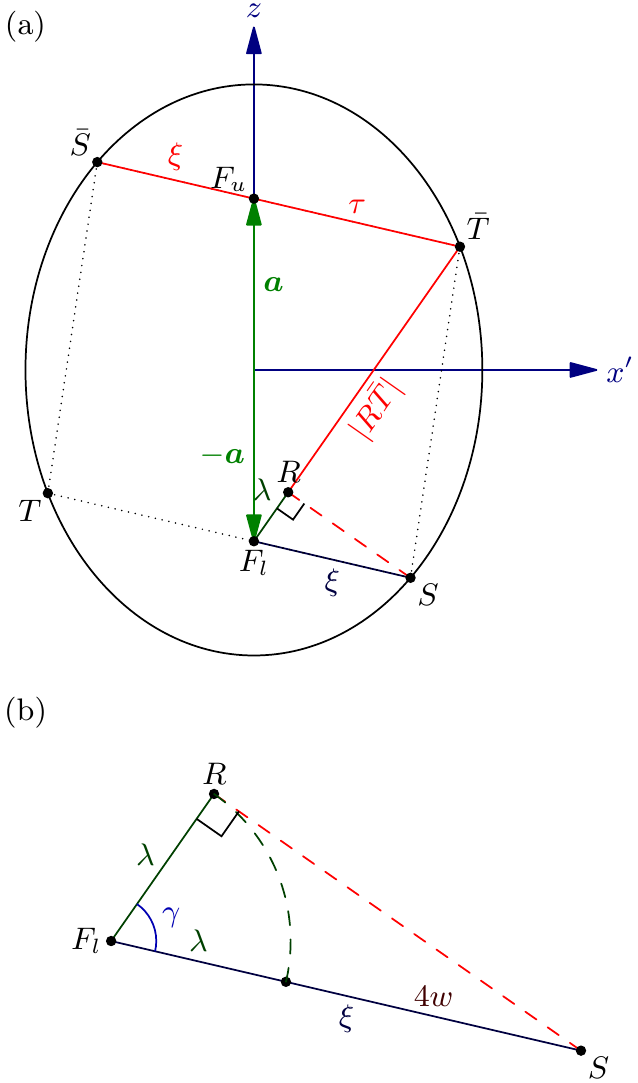}
	\caption{Geometric interpretation of the expectation value $ \mathcal{S} $ and the probability $ w $. (a) The expectation value $ \mathcal{S} $ is given by the length of the two line segments $ \abs{\bar{S}\bar{T}} $ and $ \abs{R\bar{T}} $ (solid red lines). Due to the focal point property $ \abs{F_{u}\bar{T}}+\abs{\bar{T}F_{l}}=2 $, the expectation value $ \mathcal{S} = 2 +\xi - \abs{RF_{l}} $ can be expressed by $ \xi $ and the line segment $ \lambda $ (light green line). (b) The probability $ w = (\xi -\lambda)/4$ is therefore four times the difference between the hypothenuse and the cathete of the rectangular triangle spanned by the points $ S $, $ R $ and $ F_{l} $}
	\label{fig:geo-deriv}
\end{figure}

The first scalar product
\begin{equation}\label{eq:1st-cont}
	\vec{q}\cdot (\vsk-\vtk) = \xi + \tau
\end{equation} 
is the length of the line segment $ \bar{S}\bar{T} $, since $ \vec{q} $ is parallel to $ \vsk - \vtk $. 

The measurement vector $ \vec{r} $ is in general not parallel to $ \vsk + \vtk $. Thus, the second scalar product
\begin{equation}\label{eq:2nd-cont}
	\vec{r}\cdot(\vsk+\vtk) = \abs{R\bar{T}},
\end{equation}
is the distance between the points $ \bar{T} $ and $ R $, where the latter is defined by the parallel projection of $ \vsk+\vtk $ on the vector $ \vec{r} $. 


From \cref{fig:geo-deriv}(a) and the focal point property of the ellipse, we find the relation
\begin{equation}\label{eq:absTF}
	2 - \tau = \abs{R\bar{T}} + \lambda,
\end{equation}
where $ \lambda $ is the signed length of the line segment $ \abs{RF_{l}} $ between the point $ R $ and the lower focal point $ F_{l} $. Note, that $ \lambda $ is defined positive if $ R $ is between $ \bar{T} $ and $ F_{l} $, and negative otherwise.

Inserting the expressions for the two scalar products, \cref{eq:1st-cont,eq:2nd-cont}, together with the relation between the different lengths, \cref{eq:absTF}, into the expectation value $ \mathcal{S} $, \cref{eq:S-vectors-K}, we find
\begin{equation}\label{eq:S-RF}
	\mathcal{S}= 2 + (\xi  - \lambda).
\end{equation} 
This result is understood geometrically from \cref{fig:geo-deriv}. The expectation value is the distance from the point $ \bar{S} $ via the point $ \bar{T} $ to the point $ R $. If we instead of $ R $ went to the point $ F_{l} $ this distance would be $ 2+\xi $, due to the focal point property. Since we only go to $ R $, we have to subtract the additional length $ \lambda $. We note, that $ R $ can be in principle be further away from $ \bar{T} $ than $ F_{l} $. To include this effect, $ \lambda $ is negative in this case.  

When we insert this result into \cref{eq:w-ito-S}, we find that the violation probability
\begin{equation}\label{eq:w-xi-ell}
	w=\frac{\xi-\lambda}{4}
\end{equation}
is proportional to the difference between the two lengths $ \xi $ and $ \lambda $. 
Geometrically, the violation probability manifests itself as the difference between the hypothenuse and the leg of a right triangle. This interpretation is depicted in \cref{fig:geo-deriv}(b).

In Appendix \ref{app:w-geo-tau}, we further discuss the geometric interpretation of $ w $ and derive the expression
\begin{equation}\label{eq:w-tau}
	w = \frac{\mathcal{C}^{2}}{4} \left(1- \frac{\tau \mathcal{C}^{2}}{(2-\tau)(2\tau -\mathcal{C}^{2})} \right)
\end{equation}
which only depends on the concurrence $ \mathcal{C} $ and the distance $ \tau $ between the point $ \bar{T} $ and the focal point $ F_{u} $.

\section{Optimization of the probability}\label{sec:optimization}
We now use our geometric considerations to optimize the violation probability $ w $ for a given entangled bipartite state $ \ket{\Psi} $. We start by determining the optimal violation probability for an arbitrary but fixed concurrence $ \mathcal{C} $, and then discuss the special cases of small and large concurrences. 
Moreover, we identify the optimal concurrence which optimizes the violation probability $ w $ over all concurrences, and discuss its relation to the golden ratio.

\subsection{Optimal violation probability}
We start by maximizing the probability $ w $, \cref{eq:w-tau}, for an arbitrary but fixed concurrence $ \mathcal{C} $ over all possible lengths $ \tau $ by calculating the derivative
\begin{equation}\label{key}
	\frac{\partial w}{\partial \tau} = \frac{\mathcal{C}^{4}}{2(2-\tau)(2\tau-\mathcal{C}^{2})}\left(1+\frac{\tau}{2-\tau}-\frac{2\tau}{2\tau-\mathcal{C}^{2}}\right)
\end{equation}
and determining its roots. Straightforward algebra shows, that the roots are given by
\begin{equation}\label{key}
	\tau = \pm \mathcal{C},
\end{equation}
and since $ \tau $ represents a length, only the positive solution
\begin{equation}\label{eq:tau-opt}
	\tau_{\mathrm{opt}} = \mathcal{C}.
\end{equation}
allows us to optimize the probability $ w $. 

Inserting the optimal length $ \tau_{\mathrm{opt}} $, \cref{eq:tau-opt}, into the expression of $ w $, \cref{eq:w-tau}, we obtain
\begin{equation}\label{eq:S-opt}
	w_{\mathrm{opt}}(\mathcal{C})= \mathcal{C}^{2} \frac{1-\mathcal{C}}{(2-\mathcal{C})^{2}}
\end{equation} 
as the optimal violation probability in the Hardy scenario, as a function of the concurrence $ \mathcal{C} $.

\subsection{Special cases}
We now consider two special cases of the optimal violation probability $ w_{\mathrm{opt}} $. The first case is small concurrences, that is when the common bipartite state is nearly separable. The second case we consider is the case of $ \mathcal{C} \lesssim 1 $, that is for almost maximally entangled states. 

For small concurrences, we expand the optimal expectation value, given by \cref{eq:S-opt}, into the Taylor series
\begin{equation}\label{eq:wopt-Taylor-smallC}
	w_{\mathrm{opt}} \cong \frac{\mathcal{C}^{2}}{4}+\mathcal{O}(\mathcal{C}^{4})
\end{equation}
at $ \mathcal{C}=0 $. Thus, for small entanglement, the expectation value and thus the probability $ w $ grows quadratically with the concurrence. This behavior is similar to the optimal expectation value $ \mathcal{S}_{\mathrm{opt}} = 2\sqrt{1-\mathcal{C}^{2}} $ for the CHSH scenario, which also grows quadratically in the concurrences for small $ \mathcal{C} $. In fact, the violation probability $ w_{\mathrm{opt}} $ of the Hardy scenario only differs in order $ \mathcal{O}(\mathcal{C}^{5}) $ from its corresponding value in the CHSH scenario. Therefore, for small concurrences $ \mathcal{C} $ the additional constraints \cref{eq:connection-qs,eq:connection-qt,eq:connection-rt} have a negligible effect on the achievable expectation value $ \mathcal{S}_{\mathrm{opt}} $, and thus on the optimal violation probability $ w_{\mathrm{opt}} $.

For $ \mathcal{C} \lesssim 1 $, we expand \cref{eq:S-opt} around $ \epsilon=1-\mathcal{C} $, and find
\begin{equation}\label{eq:S-opt-Taylor-epsilon}
	w_{\mathrm{opt}} \cong \epsilon-4\epsilon^{2} + \mathcal{O}(\epsilon^{3}),
\end{equation}
which decreases linearly with the concurrence when $ \mathcal{C}$ goes to $ 1 $. For maximally entangled states, that is $ \mathcal{C}=1 $, the violation probability $ w $ vanishes. This behavior differs completely from the CHSH scenario, where the optimal expectation value $ \mathcal{S}_{\mathrm{opt}} $ obtains its maximal value for the maximally entangled state, that is for $ \mathcal{C}=1 $. 

\subsection{Optimal concurrence and the golden ratio}
Since the probability $ w $ vanishes for both extreme cases $ \mathcal{C}=0 $ and $ \mathcal{C}=1 $ but is non vanishing for $ 0<\mathcal{C}<1 $, an optimal concurrence $ \mathcal{C}_{\mathrm{opt}} $ must exist, which maximizes the optimal probability $ w_{\mathrm{opt}}(\mathcal{C}) $, \cref{eq:S-opt}. 


By differentiating $ w_{\mathrm{opt}}(\mathcal{C}) $, \cref{eq:S-opt}, and determining the root of the resulting expression, the optimal concurrence reads 
\begin{equation}\label{key}
	\mathcal{C}_{\mathrm{opt}} = 3-\sqrt{5}.
\end{equation}

When we insert the optimal concurrence $ \mathcal{C}_{\mathrm{opt}} $ into the optimal length of $ \tau_{\mathrm{opt}} $, \cref{eq:tau-opt}, and
\begin{equation}\label{key}
	\xi_{\mathrm{opt}} = \frac{\mathcal{C}^{2}}{2-\mathcal{C}^{2}},
\end{equation}
we find that their ratio
\begin{equation}\label{eq:tau-xi-goldenratio}
	\frac{\tau_{\mathrm{opt}}}{\xi_{\mathrm{opt}}}= \frac{1+\sqrt{5}}{2} = \Phi,
\end{equation}
is the golden ratio $ \Phi $. Due to the relation
\begin{equation}\label{eq:2-tau-tau-goldenratio}
	\frac{\tau_{\mathrm{opt}}}{\xi_{\mathrm{opt}}} =	\frac{2-\tau_{\mathrm{opt}}}{\tau_{\mathrm{opt}}},
\end{equation}
which is valid for all concurrences $ \mathcal{C} $, the ratio between the lengths $ \tau $ and $ 2-\tau $ is also determined by the golden ratio. 

From the definition of the golden ration, the relation
\begin{equation}\label{eq:tau+xi-goldenratio}
	\frac{\tau_{\mathrm{opt}}}{\xi_{\mathrm{opt}}} = \frac{\tau_{\mathrm{opt}} + \xi_{\mathrm{opt}}}{\tau_{\mathrm{opt}}}
\end{equation}
follows directly. As a consequence, by comparing \cref{eq:2-tau-tau-goldenratio} with \cref{eq:tau+xi-goldenratio}, we find that for the optimal concurrence, we have 
\begin{equation}\label{key}
	\tau_{\mathrm{opt}} + \xi_{\mathrm{opt}} = 2-\tau_{\mathrm{opt}},
\end{equation}
that is the distance between the points $ \bar{S} $ and $ \bar{T} $, which are defined by the end points of the vectors $ -\vsk $ and $ -\vtk $ on the ellipse, is equal to the distance between the point $ \bar{T} $ and the lower focal point of the ellipse $ F_{l} $. 


\section{Optimal measurement vectors}\label{sec:optimization_meas}
In the previous section, we have calculated the optimal violation probability for any given concurrence of the underlying state. Now, we determine the associated optimal measurement vectors. Furthermore, we compare these optimal measurements to a simpler measurement strategy and discuss the effect of using non optimal measurement vectors on the violation probability.

\subsection{Optimal angles and resulting vectors}
In order to determine the optimal vectors, we first parameterize the vector 
\begin{equation}\label{eq:vecq-varphi}
	\vec{q} = \begin{pmatrix}
		\sin\theta \\ \cos\theta
	\end{pmatrix}
\end{equation}
by the polar angle $ \theta $ between the vector $ \vec{q} $ and the $ z $ axis.

\begin{figure}
	\centering
	\includegraphics{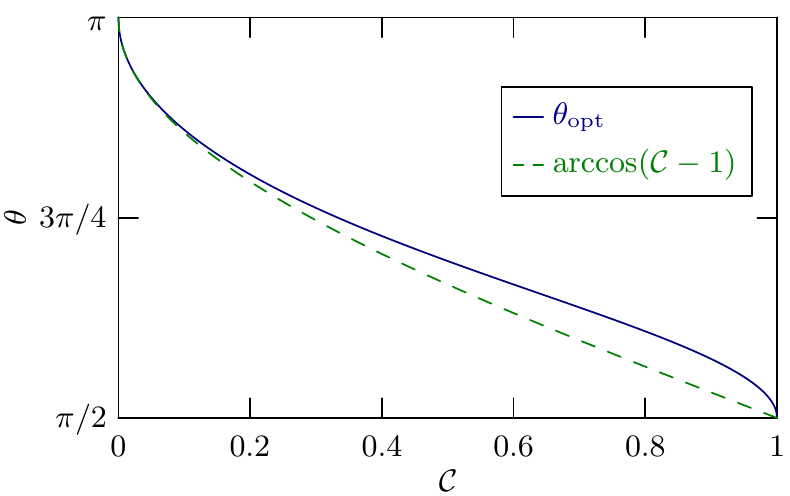}
	\caption{Optimal angle $ \theta $ between the measurement vector $ \vec{q} $ and the $ z $ axis as function of the concurrence $ \mathcal{C} $. For small concurrences $ \mathcal{C}\to 0 $ the optimal measurement vector $ \vec{q} $ points almost along the negative $ z $ axis of the coordinate system, while for large concurrences, $ \mathcal{C} \to 1 $, $ \vec{q} $ points along the $ x^{\prime} $ axis.   }
	\label{fig:optimalangle}
\end{figure}

Inserting this representation of the vector $ \vec{q} $ back into the definition of $ \tau $, \cref{eq:def-tau-ex}, and setting the result equal to the optimal length $ \tau_{\mathrm{opt}} $, \cref{eq:tau-opt}, we determine the optimal angle
\begin{equation}\label{eq:phi-opt}
	\theta_{\mathrm{opt}} = \arccos\left(-\sqrt{\frac{1-\mathcal{C}}{1+\mathcal{C}}}\right)
\end{equation}
as a function of the concurrence $ \mathcal{C} $. This function is depicted in \cref{fig:optimalangle}.

The optimal angle is therefore always in the interval $ \pi/2 $ to $ \pi $ and hence the vector $ \vec{q} $ lives in the lower half of the Bloch sphere. For $ \mathcal{C} \to 0 $ the optimal angle is close to $ \pi $, that is the vector $ \vec{q} $ points mostly along the negative $ z $ axis. With increasing concurrence $ \mathcal{C} $ the optimal angle decreases. For $ \mathcal{C} \to 1 $ the deviation from $ \vec{q} $ points mostly along the $ x^{\prime} $ axis ($\theta=\pi/2$).   

Indeed, when we insert \cref{eq:phi-opt} into \cref{eq:vecq-varphi}, we obtain
\begin{equation}\label{eq:vecq-optimal}
	\vec{q}_{\mathrm{opt}}= \frac{1}{\sqrt{1+\mathcal{C}}}\begin{pmatrix}
		\sqrt{2\mathcal{C}}\\
		-\sqrt{1-\mathcal{C}}
	\end{pmatrix},
\end{equation}
which when we also consider the azimuthal angle $ \phi $ leads to the three dimensional measurement vector
\begin{equation}\label{eq:vecq-optimal3d}
	\vec{q}_{\mathrm{opt}}= \frac{1}{\sqrt{1+\mathcal{C}}}\begin{pmatrix}
		\sqrt{2\mathcal{C}}\cos\phi\\
		\sqrt{2\mathcal{C}}\sin\phi\\
		-\sqrt{1-\mathcal{C}}
	\end{pmatrix}.
\end{equation}
When we choose $ \phi=0 $ that is the vector $ \vec{q} $ in the $ x $-$ z $ plane, inserting $ \vec{q}_{\mathrm{opt}} $, \cref{eq:vecq-optimal3d}, into the definition of $ \vec{t} $, \cref{eq:connection-qt}, leads to 
\begin{equation}\label{key}
	\vec{q}_{\mathrm{opt}} = \vec{t}_{\mathrm{opt}},
\end{equation}
and the two optimal measurement vectors point along the same direction in their respective coordinate system.  
By inserting this result into the expression for $ \vec{s} $, \cref{eq:connection-qs}, and comparing it with the expression for $ \vec{r} $, \cref{eq:connection-rt}, we furthermore find
\begin{equation}\label{key}
	\vec{r}_{\mathrm{opt}}=\vec{s}_{\mathrm{opt}} = \frac{1}{\sqrt{4-3\mathcal{C}^{2}+\mathcal{C}^{3}}}\begin{pmatrix}
		-\mathcal{C}\sqrt{2\mathcal{C}}\\
		-(2+\mathcal{C})\sqrt{1-\mathcal{C}}
	\end{pmatrix}.
\end{equation}
As a consequence, the optimal measurement setting is to choose the measurements symmetrically on both subsystems $ A $ and $ B $ of the bipartite state. 

We note that these measurement settings are assumptions explicitly made by Hardy in his original article \cite{Hardy1993}. Hence allowing $ \vec{q} $ and $ \vec{t} $ to be non equal measurement directions does not improve the probability $ w $ to violate the nonlocality assumption.  

%
%

\subsection{Nonoptimal measurements}
In the previous sections we have determined the optimal measurements. We now ask the questions of how important the exact choice of these measurement settings is in obtaining a significant violation probability. For this purpose, we investigate the case of fixing the measurement vector $ \vec{q} $ to point along the $ x^{\prime} $ axis of our coordinate system. This situation has the advantage that we do not have to adjust this measurement operator to the underlying state, but only the other three measurements $ \vec{r},\vec{s}$ and $\vec{t} $, where the latter two have the same $ z $ component and differ in the $ x^{\prime} $ component only in the sign.

When $ \vec{q} $ points along the $ x^{\prime} $ axis, it follows from the definition of the length $ \tau $, \cref{eq:def-tau}, that $ \tau = \mathcal{C}^{2} $, which by inserting into the expression for the violation probability, \cref{eq:w-tau}, leads to 
\begin{equation}\label{eq:S-r}
	w =  \frac{\mathcal{C}^{2}-\mathcal{C}^{4}}{4-2\mathcal{C}^{2}}.
\end{equation}
\begin{figure}
	\centering
	\includegraphics{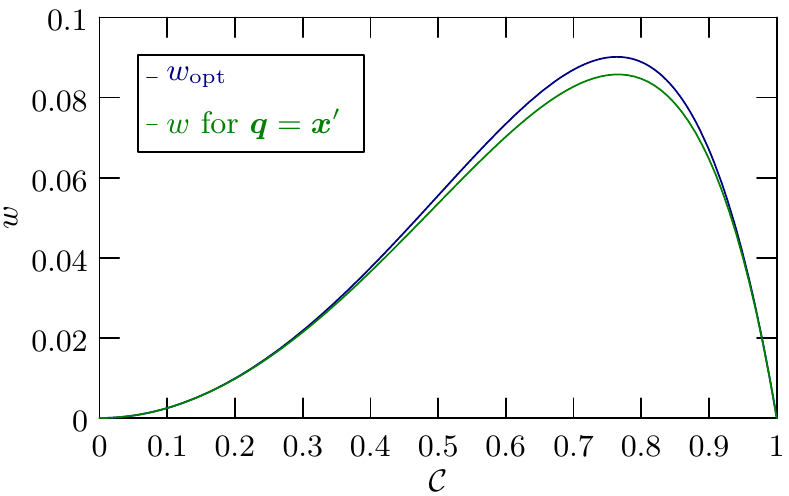}
	\caption{Violation probability $ w $ for the optimal Hardy non-locality scenario (blue curve), and the Hardy scenario with the measurement vector $ \vec{q} $ pointing along the $ \vec{x}^{\prime} $ axis (green curve). For both small and large concurrences $ \mathcal{C} $ the two curves are almost identical, and the violation probability effectively does not depend on the choice of the measurement vectors. For $ 0.5 \lesssim\mathcal{C} \lesssim 0.9 $), the optimal violation probability provides a visible improvement, up to $ \Delta w \approx 0.5\% $, over the simplified measurement strategy.}
	\label{fig:hardyvsbell}
\end{figure}
In \cref{fig:hardyvsbell}, we depict this function and compare it to the optimal measurement strategy for the Hardy scenario. 
For small concurrences, that is in the limit $ \mathcal{C}\ll 1 $, a Taylor expansion of \cref{eq:S-r} gives 
\begin{equation}\label{eq:wr-taylor-smallC}
	w \cong \frac{\mathcal{C}^{2}}{4}+\mathcal{O}(\mathcal{C}^{4}),
\end{equation}
and thus the expectation value grows quadratically in the concurrence. Compared to the Taylor expansion for small concurrences for the optimal violation probability $ w_{\mathrm{opt}} $, \cref{eq:wopt-Taylor-smallC}, the approximation $ w_{r} $ only differs up to fourth order in $ \mathcal{C} $, and thus for small concurrences the differences in the violation probability for the optimal and the case, where $ \vec{q} $ points along the $ x^{\prime} $ axis, are negligible.

This result seems quite surprising, since in the previous section, we have seen that the optimal angle for small concurrences is close to $ \theta = \pi $. Thus, for small concurrences the choice of the measurement vector $ \vec{q} $ is not the relevant parameter for increasing the achievable violation probability $ w $. 


For $ \mathcal{C} \lesssim 1 $, we perform a Taylor expansion of the violation probability, \cref{eq:S-r}, for the small parameter $ \epsilon = 1-\mathcal{C} $ 
\begin{equation}\label{eq:Sr-epsilon}
	w \cong \epsilon-\frac{9}{2}\epsilon^{2}+\mathcal{O}(\epsilon^{3}),
\end{equation}
at $ \epsilon=0 $.
The expectation value thus grows linearly with $ \epsilon $, corresponding to a linear decay towards $ 2 $ for $ \mathcal{C} \to 1 $. 
When we compare this expansion, \cref{eq:Sr-epsilon}, to the Taylor series at $ \mathcal{C}=1 $ of the optimal expectation value, \cref{eq:S-opt-Taylor-epsilon}, we find that they agree in first order and differ only slightly in the second order of $ \epsilon $. Thus, for $ \mathcal{C}\lesssim 1 $ the choice of $ \vec{q} $ pointing along the $ x^{\prime} $ axis provides a good approximation for the optimal measurement. This result is no surprise, since the optimal angle $ \theta_{\mathrm{opt}} $, \cref{eq:phi-opt}, converges to $ \pi/2 $, when $ \mathcal{C} $ goes to $ 1 $.

The difference $ \Delta w = w_{\mathrm{opt}}-w $ between the optimal violation probability and $ \vec{q} $ pointing along $ x^{\prime} $ is largest for $ \mathcal{C} \approx 0.75 $ where the violation probability is decreased by $ \Delta w \approx 0.44\% $ from the optimal probability.  

\section{Role of the constraints}\label{sec:comp-chsh}
In the previous section, we optimized the violation probability $ w $. We have demonstrated in Section \ref{sec:Hardy-CHSH} that this optimization corresponds to maximizing the expectation value $ \mathcal{S} $ familiar from the CHSH inequality under three additional constraints. In this section we study their effects. For this purpose, we consider the optimal expectation value $ \mathcal{S}_{\mathrm{opt}} $, under all possible combinations of one or two of these constraints, and compare them to the CHSH case, without any constraints, and the Hardy scenario, discussed in the previous sections.
 
\subsection{Single constraint}
We start by discussing the case of a single constraint applied to the measurement vectors. We first show that all three constraints lead to the same optimal expectation value $ \mathcal{S}_{\mathrm{opt}} $ and are therefore equivalent. 

For this purpose, we note that relabeling the subsystems $ A $ and $ B $, as well as the respective measurements does not change the physical system. 
Thus, by exchanging the vectors $ \vec{q} $ and $ \vec{t} $ as well as the vectors $ \vec{r} $ and $ \vec{s} $, the constraint between the vectors $ \vec{q} $ and $ \vec{s} $, \cref{eq:prob-vectors-2}, becomes the constraint between $ \vec{t} $ and $ \vec{r} $, \cref{eq:prob-vectors-3}, and vice versa. If we only restrict the vector $ \vec{s} $ by \cref{eq:prob-vectors-2} while we choose the vectors $ \vec{q} $, $ \vec{r} $ and $ \vec{t} $ freely, this corresponds to the same situation as restricting $ \vec{r} $ by \cref{eq:prob-vectors-3}, while choosing $ \vec{q} $, $ \vec{s} $ and $ \vec{t}$ independently. 

Furthermore, by interchanging the vector $ \vec{q} $ with $ -\vec{q} $, the restriction between $ \vec{q} $ and $ \vec{t} $, \cref{eq:prob-vectors-1}, is equivalent to the connection between the vectors $ \vec{q} $ and $ \vec{s} $, \cref{eq:prob-vectors-2}. Thus, the case of $ \vec{t} $ being determined by $ \vec{q} $, while the other measurement vectors $ \vec{q} $, $ \vec{s} $ and $ \vec{r} $ are chosen freely, is equivalent to the case of only $ \vec{s} $ being restricted by $ \vec{q} $, while $ \vec{q} $, $\vec{t}$ and $ \vec{r} $ are chosen independently. As a result, all three conditions lead to the same optimal expectation value $ \mathcal{S}_{\mathrm{opt}} $. 

\begin{figure}
	\centering
	\includegraphics{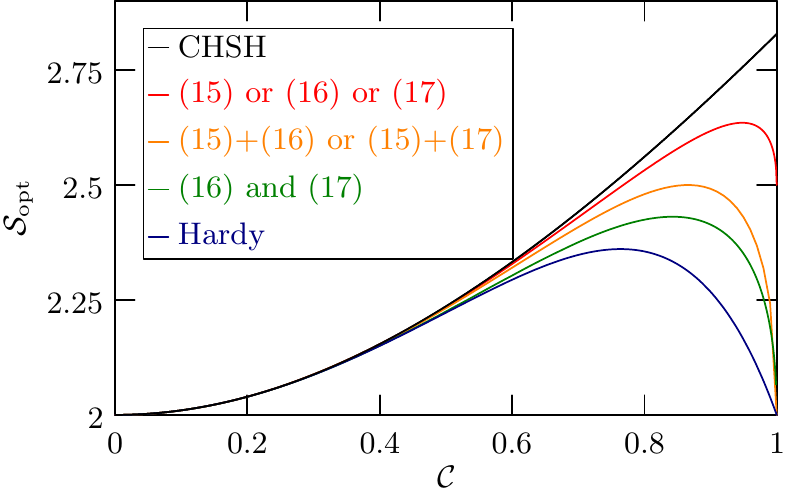}
	\caption{Dependence of the optimal expectation value $ \mathcal{S}_{\mathrm{opt}} $ on the different constraints. Without any constraints the CHSH inequality (black curve) grows monotonically with the concurrence. In the Hardy scenario (blue curve) the expectation value increases quadratically for small concurrences has an optimal concurrence of $ \mathcal{C} = 3-\sqrt{5} $. Further increasing the concurrence leads to a decrease in $ \mathcal{S}_{\mathrm{opt}} $. For two constraints (orange and green curves) the behavior is similar to the Hardy scenario, however for large concurrences the achievable expectation values are higher than for the Hardy scenario. For maximally entangled states still no violation of the CHSH inequality is achieved. If only a single constraint is valid (red curve), the expectation value again assumes its maximum for $ \mathcal{C}<1 $, however it does not decrease back to $ \mathcal{S}_{\mathrm{opt}}=2 $ for $ \mathcal{C}=1 $ as for two or all three constraints. Thus, a violation of the CHSH inequality is possible for maximally entangled states.}
	\label{fig:diffconst}
\end{figure}

Unfortunately, the optimization for only a single constraint is rather difficult to perform analytically, and we are therefore content with a numerical optimization. The optimal expectation value is shown by the red curve in \cref{fig:diffconst}. In this picture, we compare it to the CHSH case (black curve) and the Hardy scenario (blue curve).

For small concurrences $ \mathcal{C} $, the optimal expectation value is similar to both the behavior of the CHSH inequality and the Hardy scenario. This result is no surprise, since we have already demonstrated in the previous section that the Hardy scenario and the CHSH inequality lead to almost the same behavior for small concurrences, and a single condition can neither lead to a result better than the CHSH case, nor worse than the Hardy scenario. 

For medium concurrences, that is $ 0.5 \lesssim \mathcal{C} \lesssim 0.8 $, the single constraint case is still similar to the CHSH case, but starts to deviate clearly from the Hardy constraints. For large concurrences, there exists a maximum of the optimal expectation value $ \mathcal{S}_{\mathrm{opt}} \approx 2.64 $ at $ \mathcal{C}\approx 0.95 $, while for larger concurrences the optimal expectation value decreases down to $ \mathcal{S}_{\mathrm{opt}} = 2.5 $ for $ \mathcal{C}=1 $.

Hence, a single constraint is sufficient to obtain the optimal expectation value $ \mathcal{S}_{\mathrm{opt}} $ for a non maximally entangled state ($ \mathcal{C} < 1 $). The associated entanglement of the underlying state is larger than for the Hardy scenario. In contrast to the Hardy scenario, the optimal expectation value does not decrease towards $ \mathcal{S}_{\mathrm{opt}} = 2 $, that is the classical boundary, for $ \mathcal{C}\to 1 $. Thus, it is always possible to achieve a violation of the CHSH inequality with a single constraint for maximally entangled states.

\subsection{Two measurement vectors defined by a common measurement vector}
We now consider the case of two restrictions. We start by lifting the restriction between the measurement vectors $ \vec{r} $ and $ \vec{t} $, \cref{eq:prob-vectors-3}, compared to the Hardy scenario. Thus, the vectors $ \vec{s} $ and $ \vec{t} $ are still defined by the choice of $ \vec{q} $, while the vector $ \vec{r} $ is chosen independently. 

In Section \ref{sec:geometric} we demonstrated that from the two constraints, \cref{eq:prob-vectors-2,eq:prob-vectors-1}, on the vectors $ \vsk $ and $ \vtk $, it follows that the parallelogram defined by the points $ S,T,\bar{S} $ and $ \bar{T} $ intersects the $ z $ axis in the focal points $ F_{u} $ and $ F_{l} $ of the ellipse.  

Furthermore, we have seen that the measurement vector $ \vec{q} $ is parallel to the vector $ \vsk-\vtk $. The other measurement vector $ \vec{r} $ on the subsystem $ A $ is not constraint anymore, and we choose it parallel to the vector $ \vsk+\vtk $. Then, maximizing the expectation value $ \mathcal{S} $ corresponds to maximizing the perimeter of a parallelogram enclosed by an ellipse which goes through the focal points of the ellipse. 

From numerical optimization we find, that the optimal parallelogram that fulfills these conditions is the rectangle whose sides are parallel to the $ x^{\prime} $ and $ z $ axis. In this case, it is straightforward to see that the side parallel to the $ z $ axis has length $ 2\sqrt{1-\mathcal{C}^{2}} $, while the side parallel to the $ x^{\prime} $ axis is of length $ 2\mathcal{C}^{2} $. The rectangle therefore has a perimeter of $ 4(\mathcal{C}^{2}+\sqrt{1-\mathcal{C}^{2}}) $. 


Therefore, the optimal expectation value is given by
\begin{equation}\label{eq:Smax-tilde}
	\mathcal{S}_{\mathrm{opt}} = 2(\mathcal{C}^{2} + \sqrt{1-\mathcal{C}^{2}}),
\end{equation}
and consists of an increasing and a decreasing term with the concurrence $ \mathcal{C} $.

In \cref{fig:diffconst} we depict the expectation value $ \mathcal{S} $ under these constraints (orange curve), alongside the expectation values $ \mathcal{S} $ for the CHSH case (black curve), that is without any constraints, and the Hardy scenario (blue curve). 

The expectation value given by \cref{eq:Smax-tilde} is not monotonic in $ \mathcal{C} $. Instead for both $ \mathcal{C}=0 $ and $ \mathcal{C}=1 $ the maximum $ \mathcal{S}_{\mathrm{opt}} = 2 $ cannot violate the CHSH inequality, while for all concurrences in between the maximum is strictly above $ 2 $, and a violation is possible. 

\begin{figure*}
	\centering
	\includegraphics{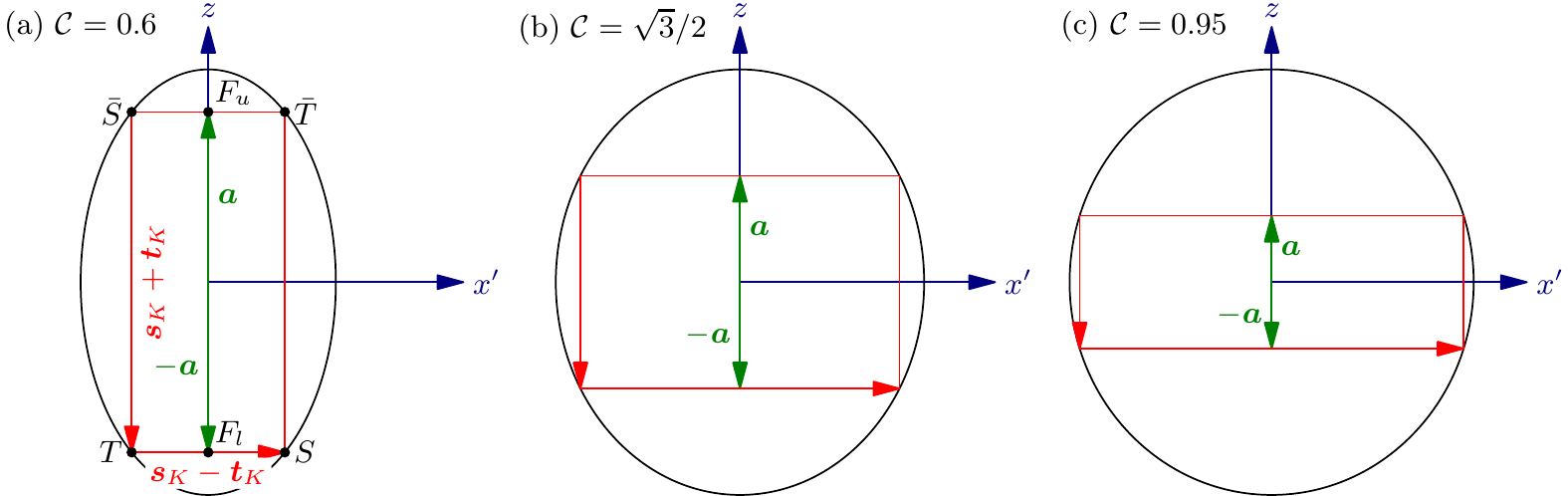}
	\caption{Rectangle with perimeter $ 2\mathcal{S}_{\mathrm{opt}} $ for three different concurrences $ \mathcal{C} $. (a) For small concurrences the focal points $ F_{u} $ and $ F_{l} $ are far apart and the rectangle is elongated along the $ z $ axis. For $ \mathcal{C} \to 0 $ the rectangle starts to collapse onto the $ z $ axis. (b) For the optimal concurrence $ \mathcal{C} = \sqrt{3}/2 $, the focal points are located exactly in the middle of the semi major axes of the ellipse, leading to a rectangle with side lengths of $ 1 $ along the $ z $ axis, and $ 1.5 $ along the $ x^{\prime} $ axis. The optimal concurrence is thus not found for a square. (c) In the case of $ \mathcal{C}\to 1$ the focal points move towards the center of the ellipse. Hence, the rectangle elongates along the $ x^{\prime} $ axis, and for $ \mathcal{C}=1 $ the rectangle degenerates into a line along the $ x^{\prime} $ axis.}
	\label{fig:rectgeo}
\end{figure*}

The decrease of $ \mathcal{S}_{\mathrm{opt}} $ for $ \mathcal{C}\to 1 $ can be understood geometrically, which is depicted in \cref{fig:rectgeo}, and is a consequence of demanding the parallelogram to go through the focal points.  Increasing $ \mathcal{C} $ increases the semi minor axis of the ellipse, and allows the side of the rectangle parallel to the semi minor axis to increase. At the same time, the focal points move in closer towards the center of the ellipse, decreasing the length of the sides parallel to the semi major axis. In the limit of approaching a maximally entangled state $ \ket{\Psi} $, the focal points collapse onto the origin of the ellipse (which is distorted to a circle) and the rectangle collapse onto a line along the $ x $ axis of our coordinate system. 
 
Finally, we note that for the concurrence $ \mathcal{C} = \sqrt{3}/2 $ the violation of the CHSH inequality is maximal, at which the expectation value, \cref{eq:Smax-tilde}, assumes the value $ \mathcal{S}_{\mathrm{opt}}=2.5 $. For larger concurrences this value rapidly decreases towards the classical bound $ \mathcal{S}_{\mathrm{opt}} = 2 $.

We conclude the situation of having two measurement vectors depending on a single measurement vector by considering the case, where only constraints between the vectors $ \vec{q} $, $\vec{r}$ and $ \vec{t} $, \cref{eq:prob-vectors-1,eq:prob-vectors-3} are present, while we lift the constraint between $ \vec{q} $ and $ \vec{s} $, \cref{eq:prob-vectors-2}. 
When we interchange the vectors $ \vec{q} $ with $ \vec{t} $ and $ \vec{r} $ with $ \vec{s} $ in the conditions, \cref{eq:prob-vectors-3,eq:prob-vectors-1}, they are identical to the conditions discussed above, \cref{eq:prob-vectors-2,eq:prob-vectors-1}. Therefore, the two cases are equivalent, and we obtain the same result for the optimal expectation value $ \mathcal{S}_{\mathrm{opt}} $.

\subsection{Two measurement vectors determined by two measurement vectors}
In contrast to the two previous cases, the case of two constraints on $ \vec{s} $ and $ \vec{r} $, \cref{eq:prob-vectors-2,eq:prob-vectors-3}, is different and cannot be mapped to the other cases of two constraints. 
We again optimize the expectation value $ \mathcal{S}_{\mathrm{opt}} $ under these constraints numerically, and depict the result by the green curve in \cref{fig:diffconst}. 

As a result, the two constraints \cref{eq:prob-vectors-2,eq:prob-vectors-3} lead to a reduced optimal expectation value $ \mathcal{S}_{\mathrm{opt}} $, compared to the case of applying the constraints, \cref{eq:prob-vectors-2,eq:prob-vectors-1} or \cref{eq:prob-vectors-1,eq:prob-vectors-3}. For small concurrences the behavior is similar to the other cases, with various constraints. For large concurrences, the optimal expectation value again decreases towards the classical limit $ \mathcal{S} =2 $ for $ \mathcal{C}\to 1 $. For $ \mathcal{C} \approx 0.84 $ we obtain the maximal expectation value $ \mathcal{S}_{\mathrm{opt}} \approx 2.43 $. Thus, the optimal concurrence is larger than for the Hardy constraints, $ \mathcal{C}_{\mathrm{opt}} = 3-\sqrt{5} $, but lower than when we apply the other two constraints simultaneously. 

We therefore conclude that two out of the three constraints for the Hardy scenario always lead to the decay of the optimal expectation value $ \mathcal{S}_{\mathrm{opt}} $ to the classical regime for maximally entangled states. The main differences between the results for these different conditions is found in the regime $ 0.5 \lesssim \mathcal{C} \lesssim 0.9 $, where the achievable expectation value crucially depends on the underlying constraints.

\section{Conclusion}\label{sec:conclusion}
In this article we have demonstrated that the Hardy scenario is equivalent to the CHSH inequality with three additional constraints. 

We have studied the influence of each individual constraint by starting from the CHSH inequality and applying the constraints individually as well as in all different combinations. As a result we find that the constraints influence the optimal concurrence for a violation significantly. A single constraint still provides a violation for maximally entangled states, while for any two combinations of constraints maximally entangled states no longer violate the CHSH inequality. 

Moreover, we developed a geometrical interpretation of the violation probability $ w $ in the Hardy scenario, and applied this picture to optimize $ w $ for all entangled pure states. We then determined the associated optimal measurement settings.
 
When we furthermore optimize the violation probability over all concurrences $ \mathcal{C} $ the value of the golden ratio appears, in complete agreement with Hardy's original article. In our approach, the golden ratio appears twice inside our geometrical picture. In both cases between line segments determined by two points on the ellipse and the focal points of this ellipse. However, a fundamental geometric explanation of why the optimum is found for this ratio remains an open question for a future publication.



\section*{Acknowledgments}
We are grateful to M. Freyberger for many fruitful discussions.
J.S. thanks the Center for Integrated Quantum Science  and  Technology  (IQ\textsuperscript{ST})  for  a  fellowship within the framework of the Quantum Alliance sponsored by the Ministry of Science, Research and Arts, Baden-W{\"u}rttemberg. T.S. acknowledges support from the EU Quantum Flagship project QRANGE (grant no. 820405). W.P.S. is grateful to Texas A{\&}M University for a Faculty Fellowship at the Hagler Institute for Advanced Study at Texas A{\&}M University and to Texas A{\&}M AgriLife Research for the support of this work. The research of IQ\textsuperscript{ST} is financially supported by the Ministry of Science, Research and Arts, Baden-W{\"u}rttemberg.

\appendix
\section{Calculation of the vectors $ \vsk$ and $\vtk $ in terms of $ \vec{q} $ and $ \vec{a} $}\label{app:parallel}
In this appendix, we express the vectors $ \vsk$, $\vtk $ and $ \vec{r} $ through the measurement vector $ \vec{q} $ and the Bloch vector $ \vec{a} $ of the state $ \ket{\Psi} $.

We first calculate the vector 
\begin{equation}\label{key}
	\vsk =  \frac{K^{2}\vec{q}-\vec{a}}{\abs{K\vec{q} -\vec{a}}},
\end{equation}
with help of the explicit expression for the correlation matrix $ K $, \cref{eq:K-diag}, and the decomposition of the vector
\begin{equation}\label{key}
	\vec{q} = \vec{q}_{\perp}+\vec{q}_{\parallel}
\end{equation}
into a part that is orthogonal ($\vec{q}_{\perp}$) and one that is parallel to the Bloch vector $ \vec{a} $, as
\begin{equation}\label{key}
	\vsk =  \frac{\mathcal{C}^{2}\vec{q}_{\perp}+\vec{q}_{\parallel}-\vec{a}}{\abs{K\vec{q} -\vec{a}}},
\end{equation}
which we rearrange to 
\begin{equation}\label{eq:app-vsk-rearranged}
	\vsk =  \frac{\mathcal{C}^{2}\vec{q}}{\abs{K\vec{q} -\vec{a}}} + \frac{(1-\mathcal{C}^{2})\vec{q}_{\parallel}-\vec{a}}{\abs{K\vec{q} -\vec{a}}}.
\end{equation}
The first term on the right hand side therefore points along the direction $ \vec{q} $, while the second term points along the Bloch vector $ \vec{a} $, since both $ \vec{q}_{\parallel} $ and $ \vec{a} $ point along the same direction.

From the definition of the Bloch vector, \cref{eq:veca}, we find
\begin{equation}\label{eq:app-numerator-final}
	(1-\mathcal{C}^{2})\vec{q}_{\parallel}-\vec{a} = - \left(1-\sqrt{1-\mathcal{C}^{2}}q_{\parallel}\right)\vec{a},
\end{equation}
where $ q_{\parallel}$ is the component of $ \vec{q} $ parallel to the Bloch vector. 

As a last step we still have to evaluate the absolute value 
\begin{equation}\label{key}
	\abs{K\vec{q}-\vec{a}} = \sqrt{\vec{q}^{T}K^{2}\vec{q}+\abssq{\vec{a}}-2\vec{a}^{T}K\vec{q}}.
\end{equation} 
By inserting the definitions of the correlation matrix $ K $, \cref{eq:K-diag}, and the Bloch vector $\vec{a}$, \cref{eq:veca}, as well as using the decomposition of the vector $ \vec{q} $ into an orthogonal ($ q_{\perp} $) and parallel($ q_{\parallel} $) component with respect to the Bloch vector, we obtain 
\begin{equation}\label{key}
	\abs{K\vec{q}-\vec{a}} = \sqrt{\mathcal{C}^{2}{q}^{2}_{\perp}+{q}_{\parallel}^{2}+1-\mathcal{C}^{2}-2\sqrt{1-\mathcal{C}^{2}}{q}_{\parallel}}.
\end{equation} 

Since the measurement vector $ \vec{q} $ is a unit vector, it fulfills the relation $ q_{\perp}^{2}+q_{\parallel}^{2} = 1 $, leading to 
\begin{equation}\label{key}
	\abs{K\vec{q}-\vec{a}} = \sqrt{1-2\sqrt{1-\mathcal{C}^{2}}{q}_{\parallel}+ (1-\mathcal{C}^{2})q_{\parallel}^{2}},
\end{equation} 
which finally simplifies to 
\begin{equation}\label{eq:app-denominator-final}
	\abs{K\vec{q}-\vec{a}} = 1-\sqrt{1-\mathcal{C}^{2}}{q}_{\parallel}.
\end{equation} 

When we insert \cref{eq:app-denominator-final} together with \cref{eq:app-numerator-final} back into the expression for the vector $ \vsk $, \cref{eq:app-vsk-rearranged}, we obtain
\begin{equation}\label{key}
	\vsk = \xi \vec{q} - \vec{a},
\end{equation}
where we defined the length
\begin{equation}\label{key}
	\xi \equiv \frac{\mathcal{C}^{2}}{\abs{K\vec{q}-\vec{a}}} = \frac{\mathcal{C}^{2}}{1-\sqrt{1-\mathcal{C}^{2}}q_{\parallel}}
\end{equation}
of the component of $ \vsk $ along the vector $ \vec{q} $.

In complete analogy to the above decomposition of the vector $ \vsk $, the vector 
\begin{equation}\label{key}
	\vtk = - \frac{K^{2}\vec{q}+\vec{a}}{\abs{K\vec{q} +\vec{a}}},
\end{equation}
is decomposed into
\begin{equation}\label{eq:app:def-tau}
	\vtk = -\tau\vec{q}-\vec{a},
\end{equation}
with the length
\begin{equation}\label{key}
	\tau \equiv \frac{\mathcal{C}^{2}}{\abs{K\vec{q}+\vec{a}}}= \frac{\mathcal{C}^{2}}{1+\sqrt{1-\mathcal{C}^{2}}q_{\parallel}}.
\end{equation}

We finally determine the vector $ \vec{r} $ in terms of the vectors $ \vec{q} $ and $ \vec{a} $. With help of its definition, \cref{eq:connection-rt}, and the explicit expression for the vector $ \vtk $, \cref{eq:app:def-tau}, we rewrite the vector
\begin{equation}\label{eq:q-r:1}
	\vec{r} =  \frac{\tau\vec{q}-2\vec{a}}{\abs{\vtk-\vec{a}}},
\end{equation}
in terms of the measurement vector $ \vec{q} $, the Bloch vector $ \vec{a} $ and the lengths $ \tau $ and $ \abs{\vtk - \vec{a}} $. 

From the focal point property of the ellipse, which constitutes that the distance from one focal point via any point on the ellipse to the other focal point is twice the length of the semimajor axis, and our geometrical picture, \cref{fig:geo-int}, it immediately follows that
\begin{equation}\label{eq:tk-a.tau}
	\abs{\vtk-\vec{a}} = 2 -\tau, 
\end{equation} 
and thus the vector $ \vec{r} $, \cref{eq:q-r:1}, simplifies to
\begin{equation}\label{eq:app:q-r}
	\vec{r} = \vec{q} - \frac{2}{2-\tau}\left(\vec{q}+\vec{a}\right).
\end{equation}

\section{Parametrizing $\mathcal{S}$ in terms of the polar angle $ \theta $}\label{app:paraStheta}
The expectation value $ \mathcal{S} $ for the Hardy scenario depends only on a single measurement vector. In this appendix, we parametrize this vector in spherical coordinates and derive the resulting equation for $ \mathcal{S} $ in terms of these coordinates. 

We start by writing the measurement vector 
\begin{equation}\label{key}
	\vec{q} =\begin{pmatrix}
		\cos\phi \sin\theta \\ \sin\phi \sin\theta \\ \cos \theta
	\end{pmatrix} 
\end{equation}
in the conventional spherical coordinates, where $ \theta $ is the polar angle, and $ \phi $ denotes the azimuthal angle. 

From the connection between the vectors $ \vec{s} $ and $ \vec{q} $, \cref{eq:connection-qs}, we find 
\begin{equation}\label{key}
	\vec{s} = \frac{1}{1-\sqrt{1-\mathcal{C}^{2}}\cos\theta}\begin{pmatrix}
		-\mathcal{C}\cos\phi \sin\theta \\ \mathcal{C}\sin\phi \sin\theta \\ \cos \theta - \sqrt{1-\mathcal{C}^{2}}
	\end{pmatrix} 
\end{equation}
in these coordinates, where we further made use of the definitions of the correlation matrix $ K $, \cref{eq:K-diag}, and the Bloch vector $ \vec{a} $, \cref{eq:veca}, in terms of the concurrence $ \mathcal{C} $. 

Analogously, we find the measurement vector
\begin{equation}\label{eq:app:vect-spherical}
	\vec{t} = -\frac{1}{1-\sqrt{1+\mathcal{C}^{2}}\cos\theta}\begin{pmatrix}
		-\mathcal{C}\cos\phi \sin\theta \\ \mathcal{C}\sin\phi \sin\theta \\ \cos \theta + \sqrt{1-\mathcal{C}^{2}}
	\end{pmatrix} 
\end{equation}
from the connection between $ \vec{t} $ and $ \vec{q} $, \cref{eq:connection-qt}. 

When we insert \cref{eq:app:vect-spherical} into the relation between the vectors $ \vec{r} $ and $ \vec{t} $, \cref{eq:connection-rt}, we finally obtain the vector
\begin{widetext}
	\begin{equation}\label{key}
		\vec{r} = -\frac{1}{\sqrt{\mathcal{C}^{2}\sin^{2}\theta+((2-\mathcal{C}^{2})\cos\theta+2\sqrt{1+\mathcal{C}^{2}})^{2}}}\begin{pmatrix}
			\mathcal{C}^{2}\cos\phi \sin\theta \\ \mathcal{C}^{2}\sin\phi \sin\theta \\ (2-\mathcal{C}^{2})\cos \theta + 2\sqrt{1-\mathcal{C}^{2}}
		\end{pmatrix} .
	\end{equation}

We now have determined all the measurement vectors in spherical coordinates. Inserting these expressions into the definition of the expectation value, \cref{eq:S-avg-vectors}, and making use of the correlation matrix $ K $, \cref{eq:K-diag}, we finally obtain
\begin{equation}\label{key}
	\mathcal{S} = \frac{2}{\mathcal{C}^{2}\cos^{2}\theta+\sin^{2}\theta}\left(\mathcal{C}^{2}+\frac{2\sqrt{1-\mathcal{C}^{2}}\sin^{2}\theta\left((2-\mathcal{C}^{2}-\mathcal{C}^{4})\cos\theta+2\sqrt{1-\mathcal{C}^{2}} \right)}{\sqrt{\mathcal{C}^{2}\sin^{2}\theta+\left((2-\mathcal{C}^{2})\cos\theta+2\sqrt{1+\mathcal{C}^{2}}\right)^{2}}}\right),
\end{equation}
\end{widetext}
which is only depended on the polar angle $ \theta $, but not on the azimuthal angle $ \phi $.

\section{Derivation of $ w $ as a function of $ \tau $}\label{app:w-geo-tau}
In this appendix, we derive the violation probability, which according to \cref{eq:w-xi-ell}, is given by
\begin{equation}\label{key}
	w = \frac{\xi-\lambda}{4}
\end{equation}
with the lengths $ \xi $ and $ \lambda $, as a function of the length $ \tau $ and the concurrence $ \mathcal{C} $.
From our geometrical picture, \cref{fig:geo-deriv}, we can deduce that $ \ell $ is the leg of a right triangle with hypotenuse of length $ \xi $. 
When can therefore express the length of the leg
\begin{equation}\label{eq:RF-gamma}
	\lambda = \xi \cos\gamma
\end{equation}
in terms of the length $ \xi $ and the inner angle $ \gamma $ between this leg and the hypotenuse.

Inserting this result into \cref{eq:w-xi-ell}, the probability
\begin{equation}\label{eq:w-xi-gamma}
	w= \frac{\xi}{4}\left(1-\cos\gamma\right)
\end{equation}
is completely determined by the length $ \xi $ of the hypotenuse and the inner angle $ \gamma $ of the triangle $ SRF_{l} $. 


In order determine this angle $ \gamma $, we note from \cref{fig:geo-deriv} that the angle $ \gamma $ is also an inner angle of the triangle $ F_{u}\bar{T}F_{l} $. Since we know all the side lengths of this triangle, we immediately exploit the cosine rule and obtain 
\begin{equation}\label{eq:cosgamma-1}
	\cos\gamma = \frac{2\mathcal{C}^{2}}{(2-\tau)\tau} - 1.
\end{equation}

When we insert this result back into our expression for the probability $ w $, \cref{eq:w-xi-gamma}, the probability reads
\begin{equation}\label{eq:S-tau-xi}
	w =  \frac{\xi}{2} \left(1- \frac{\mathcal{C}^{2}}{(2-\tau)\tau} \right)
\end{equation}
as a function of the lengths $ \xi $ and $ \tau $.


Finally, from their definitions, \cref{eq:def-tau,eq:def-xi}, we relate the two lengths $ \tau $ and $ \xi $ by
\begin{equation}\label{eq:xi+tau-relation}
	\xi+\tau = \frac{2}{\mathcal{C}^{2}}\xi\tau,
\end{equation}
which allows us to rewrite the probability $ w $, \cref{eq:S-tau-xi}, as
\begin{equation}\label{eq:app:w-tau}
	w = \frac{\mathcal{C}^{2}}{4} \left(1- \frac{\tau \mathcal{C}^{2}}{(2-\tau)(2\tau -\mathcal{C}^{2})} \right),
\end{equation}
which now only depends on the length $ \tau $. 

We furthermore notice, that in order to obtain a more geometric interpretation of the violation probability $ w $, we rewrite \cref{eq:S-tau-xi} with help of \cref{eq:xi+tau-relation} as
\begin{equation}\label{eq:S-xi/2tau}
	w =  \frac{\mathcal{C}^{2}}{4} \left(1- \frac{\xi}{2-\tau} \right).
\end{equation}
For a fixed concurrence $ \mathcal{C} $, the violation probability $ w $ thus only depends on the ratio between the lengths $ \xi $ of the line segment $ SF_{l} $ and $ 2-\tau $ of the line segment $ \bar{T}F_{l} $. In order to maximize the violation probability for a fixed concurrence, we thus have to minimize the ratio between the lengths $ \xi $ and $ 2 - \tau $.

	\end{document}